\documentclass{SciPost}

\hypersetup{
    colorlinks,
    linkcolor={red!50!black},
    citecolor={blue!50!black},
    urlcolor={blue!80!black}
}

\usepackage[bitstream-charter]{mathdesign}
\DeclareSymbolFont{usualmathcal}{OMS}{cmsy}{m}{n}
\DeclareSymbolFontAlphabet{\mathcal}{usualmathcal}

\fancypagestyle{SPstyle}{
\fancyhf{}
\lhead{\colorbox{scipostblue}{\bf \color{white} ~SciPost Physics }}
\rhead{{\bf \color{scipostdeepblue} ~Submission }}

\fancyfoot[C]{\textbf{\thepage}}
}

\usepackage{color}
\usepackage{latexsym}
\usepackage{amsmath, mathdots, amscd}
\usepackage{multirow}
\usepackage{graphicx}
\usepackage{euscript}
\usepackage{verbatim}
\usepackage{epsfig}
\usepackage{hyperref}

\usepackage{float}
\usepackage{cite}
\usepackage{mathtools}
\usepackage{framed}
\def\be{\begin{eqnarray}}
\def\ee{\end{eqnarray}}
\def\0{\nonumber}

\newtheorem{Theorem}{Theorem}[section]

\newtheorem{Corollary}[Theorem]{Corollary}

\newtheorem{conj}{Conjecture} [section]

\providecommand{\vev}[1]{\langle#1\rangle}

\def\0{\nonumber}

\def\c{{\bf c}}

\usepackage{makeidx}
\usepackage{cleveref}

\begin{document}

\pagestyle{SPstyle}

\begin{center}{\Large \textbf{\color{scipostdeepblue}{
Duality-Symmetry Enhancement in Maxwell Theory\\
}}}\end{center}

\begin{center}\textbf{
Shani Meynet\textsuperscript{1,2 $\star$},
Daniele Migliorati\textsuperscript{1,2 $\dagger$},
Raffaele Savelli\textsuperscript{3 $\ddagger$}, and
Michele Tortora\textsuperscript{4,5 $\circ$}
}\end{center}

\begin{center}
{\bf 1} Mathematics Institute, Uppsala University, Box 480, SE-75106 Uppsala, Sweden
\\
{\bf 2} Center For Geometry and Physics, Uppsala University, Box 480, SE-75106 Uppsala, Sweden
\\
{\bf 3} Dipartimento di Fisica \& INFN, Universit\`a di Roma ``Tor Vergata'', \\ Via della Ricerca
Scientifica 1, I-00133 Roma, Italy
\\
{\bf 4} SISSA, Via Bonomea 265, 34136 Trieste, Italy
\\
{\bf 5} INFN, Sezione di Trieste, Via Valerio 2, 34127 Trieste, Italy
\\[\baselineskip]
$\star$ \href{mailto:shani.meynet@math.uu.se}{\small shani.meynet@math.uu.se}\,,\quad
$\dagger$ \href{mailto:daniele.migliorati@math.uu.se}{\small daniele.migliorati@math.uu.se}\,,\\
$\ddagger$ \href{mailto:raffaele.savelli@roma2.infn.it}{\small raffaele.savelli@roma2.infn.it}\,,\quad
$\circ$ \href{mailto:michele.tortora@sissa.it}{\small michele.tortora@sissa.it}
\end{center}

\section*{\color{scipostdeepblue}{Abstract}}
\textbf{\boldmath{%
Free Maxwell theory on general four-manifolds may, under certain conditions on the background geometry, exhibit holomorphic factorization in its partition function. We show that when this occurs, new discrete symmetries emerge at orbifold points of the conformal manifold. These symmetries, which act only on a sublattice of flux configurations, are not associated with standard dualities, yet they may carry 't Hooft anomalies, potentially causing the partition function to vanish even in the absence of apparent pathologies. We further explore their non-invertible extensions and argue that their anomalies can account for zeros of the partition function at smooth points in the moduli space.
}}

\vspace{\baselineskip}




\vspace{10pt}
\noindent\rule{\textwidth}{1pt}
\tableofcontents
\noindent\rule{\textwidth}{1pt}
\vspace{10pt}


\section{Introduction}

The way symmetries are understood in Quantum Field Theory has undergone a profound shift after the seminal work \cite{Gaiotto:2014kfa}. In the new paradigm of topological operators, symmetries admit a much richer variety of manifestations, whose formulation goes far beyond the realm of Group Theory and affects different areas of Physics. This generalization has also involved the closely related concept of dualities, which, as opposed to symmetries, connect different theories (or, as is relevant here, a single theory at different parameter values) that often cannot be described in terms of mutually local fields.\footnote{Recent reviews on the subject include \cite{Cordova:2022ruw,Schafer-Nameki:2023jdn,Brennan:2023mmt,Bhardwaj:2023kri,Luo:2023ive,Shao:2023gho,Costa:2024wks}.}

The quantum-mechanical behavior of the theory under symmetry and duality transformations may be plagued by 't Hooft anomalies \cite{tHooft:1979rat}, which are failures of the partition function to be a gauge-invariant function of the corresponding background fields. Such anomalies, which can also mix different symmetries, are notoriously invaluable characteristics of the theory, allowing to relate opposite regimes of it, due to their invariance across energy scales. In particular, the so-called mixed 't Hooft anomalies may be thought of as obstructions to the simultaneous gauging of the symmetries that are being mixed.

All of the above ingredients will feature prominently in this paper, which addresses them in the special case of Maxwell theory in four dimensions. Despite the absence of local interactions, when placed on topologically non-trivial  four-manifolds, this theory exhibits a wealth of intricate phenomena. In particular, as shown by \cite{Witten:1995gf,Verlinde:1995mz,Alvarez:1999uq} (and further explored in \cite{Seiberg:2018ntt}), one of its distinctive properties, i.e.~electromagnetic duality, can conflict with the classical gravitational background. Specifically, under a duality transformation, the partition function of the theory acquires an anomalous multiplicative factor that cannot be removed by a suitable choice of local counterterms. Such an occurrence has more drastic consequences when we focus on a self-dual theory, whereby (some of) the duality transformations become actual symmetries of that theory: In this case, the anomaly forces its partition function to vanish.

Zero loci of a (quasi-)modular function, such as the Maxwell partition function in terms of the (complexified) gauge coupling, are arguably much easier to determine than anomalous phases. This naturally leads us to wonder whether the zeros of the Maxwell partition function \emph{always} signal the presence of an anomalous duality-symmetry. One of the central aims of this paper is to investigate this question and prove that, in some cases, the answer is affirmative. We show that, whenever certain specific topological and geometrical conditions on the gravitational background are met, allowing the partition function to have a factorized holomorphic component, new symmetries arise at self-dual values of the coupling, which, however, are not inherited from dualities. We dub them ``partial'' symmetries because, as we will explain, they only act non-trivially on a sublattice of flux configurations (e.g.~self-dual fluxes). Their mixed 't Hooft anomaly with gravity explains why the partition function can vanish even if the ordinary duality-symmetries are all anomaly free. We will describe the properties of these new symmetries and explore their action on the extended operators of the theory, such as lines and one-form symmetry defects. Hence, to summarize, we prove that, if the gravitational background is such that the partition function admits a holomorphic factor which vanishes at a self-dual value of the coupling, then the theory enjoys a global symmetry plagued by a mixed 't Hooft anomaly with gravity.

As mentioned earlier, within the more recent research wave on Generalized Symmetries, electromagnetic duality has also been generalized and, more precisely, promoted to a non-invertible duality. This generalization employs the gauging of a discrete non-anomalous subgroup of the electromagnetic one-form symmetry of the theory \cite{Choi:2021kmx,Kaidi:2021xfk,Choi:2022zal,Cordova:2023ent}, and extends the group of duality transformations acting on the coupling to $SL(2,\mathbb{Q})$ \cite{Niro:2022ctq,Sela:2024okz}, or even to the whole $SL(2,\mathbb{R})$ as shown in \cite{Hasan:2024aow,Paznokas:2025epc}. As a consequence, now any value of the coupling is fixed by a subgroup of the non-invertible duality and thus corresponds to a self-dual theory. There is, however, to the best of our knowledge, no direct method to assess whether the corresponding non-invertible duality-symmetry suffers from a mixed 't Hooft anomaly with gravity because no framework is available to derive (or even to define) the anomalous phase of the partition function.

It is here that the alternative strategy of looking for vanishing loci of the partition function to detect anomalies mostly pays off.
We study, in particular, certain favorable situations where we control the zeros of the Maxwell partition function, namely simply-connected four-manifolds with a specific class of lattices of integral harmonic two-forms. We construct them via connected sums of more elementary four-manifolds. In each of these cases, the relevant part of the partition function has the form of a well-known generalized Jacobi theta function, having a simple zero at a \emph{smooth} point in the space of inequivalent couplings. Hence, the theory corresponding to that value of the coupling has a vanishing partition function, despite the fact that all of its invertible symmetries are free of anomalies. Armed with the piece of evidence that we built in the invertible case, we conjecture that the pathological behavior of such theories is due to a mixed anomaly between gravity and the non-invertible duality-symmetries (or possibly the partial version thereof) which emerge in this context.

The structure of this paper is as follows. In Section \ref{Sec:Inv}, after setting up the stage for our investigation by revisiting Witten's work on electromagnetic duality and reviewing the necessary mathematical background, we introduce the novel notion of partial symmetries and discuss the physical meaning of the conditions that allow them to emerge. In Section \ref{Sec:Non-Inv}, we turn our attention to the recent non-invertible generalization of electromagnetic duality, and leverage results from the previous section to probe anomalies of non-invertible symmetries through the zeros of the partition function. Finally, in Section \ref{Sec:Conclu}, we draw our conclusions and give some hints for further exploration. Appendix \ref{MathToolkit} contains a concise mathematical toolkit that helps the unfamiliar reader navigate the numerous definitions and theorems recalled in the main text.

\section{The invertible case}\label{Sec:Inv}

This section deals with invertible duality-symmetries of Maxwell theory on general four-manifolds. We start in Section \ref{wittenWork} by reviewing Witten's work on the duality properties of the Maxwell partition function. In Section \ref{SymAnom}, we discuss the anomalous phase induced by duality transformations in the presence of non-trivial gravitational backgrounds, and study its relationship with the vanishing of the partition function. In particular, we focus on the cases in which the latter splits into multiple factors, whereby new symmetries arise at special values of the coupling. Finally, in Section \ref{ActLines}, we analyze the effect of these new symmetries on the extended operators of the theory.

\subsection{Revisiting Witten's work}\label{wittenWork}

Consider Maxwell theory on a closed, connected, and oriented four-manifold $\mathcal{M}$, free of singularities, equipped with a smooth metric $g$ of Euclidean signature, and whose middle cohomology is absent of torsion. This theory has a complex gauge coupling 
\begin{equation}
	\tau=\frac{\theta}{2\pi}+\frac{2\pi{\rm i}}{e^2}\,,
\end{equation}
living in the upper-half plane $\mathbb{H}$, in terms of which the classical action reads\footnote{If $\mathcal{M}$ is non-spin, there can also be a coupling of the form $ {\rm i}\pi\int_{\mathcal{M}}c_1\cup w_2$, with $c_1$ the first Chern class of the Maxwell line bundle and $w_2$ the second Stiefel-Whitney class of $\mathcal{M}$. Although this term would slightly modify the modular behavior of the partition function, it would not change the main conclusions of this paper. Therefore, we will simply ignore it throughout.}
\be\label{MaxwellAction}
S(\tau,F)&=&\int_{\mathcal{M}} \frac{1}{2 e^2}\,F\wedge \star
F + \frac{{\rm i}\,\theta}{8\pi^2}\,F\wedge F \nonumber \\ \nonumber\\ 
&=& \frac{{\rm i}}{4\pi}\int_{\mathcal{M}}\bar{\tau}\,F^+\wedge F^++\tau\, F^-\wedge F^-\,.
\ee
Here $F$ denotes the electromagnetic field strength, $F^{\pm}=\tfrac12(F\pm \star F)$ its self-dual and antiself-dual parts, and $\star$ the (metric-dependent) Hodge-star operation.~$F$ represents a cohomology class in the image of the map $H^2(\mathcal{M},2\pi\mathbb{Z}) \xhookrightarrow{} H^2(\mathcal{M},\mathbb{R})$, and can be written locally in terms of a gauge potential as $F\stackrel{{\rm loc}}{=}{\rm d}A$.

One of the distinctive features of this theory is the celebrated electromagnetic (or $S$) duality in the space of couplings, that is, a correspondence between the quantum theory with coupling $\tau$ and the one with coupling $-1/\tau$ \cite{Witten:1995gf,Verlinde:1995mz}.\footnote{The first observations date back to \cite{Cardy:1981qy,Cardy:1981fd,Shapere:1988zv}.} The precise formulation of it uses the partition function $Z[g,\tau]$, which we take as a definition of the quantum theory. It reads
\be\label{PFMaxwell}
Z[g,\tau]&=&({\rm Im}\,\tau)^{\tfrac12(B_1-B_0)}\int \mathcal{D}A \,e^{-S(\tau,F)}\,,
\ee
where the $\tau$-dependent prefactor is a local counterterm acting as a UV regulator through a lattice with a $B_k$-dimensional space of $k$-forms.

$S$-duality, at the core, really turns out to amount to a Poisson resummation on $Z[g,\tau]$; however, it also comes with certain subtle determinants, which, for four-manifolds with non-trivial topologies, do not leave $Z$ invariant. In what follows, we will review Witten's computation \cite{Witten:1995gf} leading to this claim, but retaining \emph{all} factors, including the $\tau$-independent ones, which have been discarded in \cite{Witten:1995gf}. This will allow us to derive a more general duality-transformation law for the partition function, which correctly reproduces Witten's result when $\mathcal{M}$ is a spin manifold.

Unlike ordinary symmetries, $S$-duality exchanges mutually non-local degrees of freedom. This means that, to make it manifest, we have to extend the action \eqref{MaxwellAction} to include the dual field $\tilde{A}$, with the caveat that $A$ and $\tilde{A}$ cannot be treated as dynamical variables simultaneously. One way to implement this is to introduce two real-valued two-form fields $\mathcal{F}$ and $\tilde{F}$, coupled according to the following action
\be\label{ExtMaxAct}
\mathcal{S}&=&\frac{{\rm i}}{2\pi}\int_{\mathcal{M}}\tilde{F}\wedge (\mathcal{F}-{\rm d}A)+S(\tau,\mathcal{F})\,,
\ee
where by $``{\rm d}A"$  we mean a closed, integrally-quantized two-form, and the last piece is the action \eqref{MaxwellAction} written in terms of $\mathcal{F}$. To quantize $\mathcal{S}$, we are instructed to path-integrate over $\tilde{F},\mathcal{F}, A$. From \eqref{ExtMaxAct} it is clear that both $\tilde{F}$ and $A$ act as Lagrange multipliers. Integrating out the former freezes $\mathcal{F}$ to $\mathcal{F}\stackrel{{\rm loc}}{=}{\rm d}A$ and we get back the electric frame, i.e.~the quantum theory corresponding to \eqref{MaxwellAction}. If instead we integrate out $A$, $\tilde{F}$ localizes on closed two-forms with integral periods, and the path integral on its local potential $\tilde{A}$ yields the magnetic-dual frame of the same theory. 

Let us spell out in more detail the second manipulation. If the path integral is appropriately normalized with the volume of the gauge group, after path-integrating over $A$ we get the following expression\footnote{In order not to clutter the notation, we suppress the dependence of $Z$ on the space-time metric $g$.}
\be\label{ExtZ}
Z[\tau]&=&({\rm Im}\,\tau)^{\frac{1}{2}(B_1-B_0)}\int \mathcal{D}\mathcal{F} \,\mathcal{D}\tilde{A}\,e^{-\mathcal{\tilde{S}}}\,,
\ee
where
\be\label{DualMaxAct}
\mathcal{\tilde{S}}&=&\frac{{\rm i}}{4\pi}\int_{\mathcal{M}} \bar{\tau}\,\mathcal{F}^+\wedge \mathcal{F}^++\tau\, \mathcal{F}^-\wedge \mathcal{F}^-+2\,\tilde{F}^+\wedge \mathcal{F}^++2\,\tilde{F}^-\wedge \mathcal{F}^-\nonumber\\ \nonumber\\ &=& S(1,\mathcal{G})+S(-\tfrac{1}{\tau},\tilde{F})\,
\ee
In the second equality above we have completed the square and defined the real-valued two-form $\mathcal{G}$, whose self-dual and antiself-dual parts are
\be
\mathcal{G}^+=\sqrt{\bar\tau} \,\mathcal{F}^++\frac{1}{\sqrt{\bar\tau}}\,\tilde{F}^+\,,\qquad \mathcal{G}^-=\sqrt{\tau} \,\mathcal{F}^-+\frac{1}{\sqrt{\tau}}\,\tilde{F}^-\,,
\ee
whereas the last piece in \eqref{DualMaxAct} is the magnetic-dual Maxwell action, featuring the dual field strength and inverse gauge coupling. At this point, we can convert the path integral on $\mathcal{F}$ into a $\tau$-independent path integral on $\mathcal{G}$, storing the $\tau$ dependence in the Jacobian factor:
\be
\int \mathcal{D}\mathcal{F}\,e^{-S(1,\mathcal{G}(\mathcal{F}))} =\bar\tau^{-\frac{B_2^+}{2}}\,\tau^{-\frac{B_2^-}{2}}\,\int\mathcal{D}\mathcal{G}\,e^{-S(1,\mathcal{G})}=({\rm i}\bar\tau)^{-\frac{B_2^+}{2}}\,(-{\rm i}\tau)^{-\frac{B_2^-}{2}}\,,
\ee
where, to evaluate the gaussian path integral on $\mathcal{G}$, we have used an appropriately-normalized basis of (anti)self-dual two-forms:
\be
\mathcal{G}^\pm=\sum_{k=1}^{B_2^\pm} 2\pi\, \mathcal{G}_k^\pm\rho^\pm_k\,,\qquad{\rm with}\qquad \int_\mathcal{M}\rho_k^\pm\wedge\rho_l^\pm=\pm\delta_{kl}\qquad {\rm and}\qquad \mathcal{G}^\pm_k\in\mathbb{R}\quad \forall k\,.
\ee
Now, recalling that $B_2^\pm-B_1+B_0=\tfrac12(\chi\pm\sigma)$ are finite quantities in the continuum limit, with $\chi,\sigma$ being the Euler characteristic and the signature of $\mathcal{M}$ respectively, Eq.~\eqref{ExtZ} becomes
\be\label{Sduality}
Z[-\tfrac1\tau]&=&e^{\frac{\pi{\rm i}\sigma}{4}}\,\tau^{\frac{\chi-\sigma}{4}}\,\bar\tau^{\frac{\chi+\sigma}{4}}\,Z[\tau]\,.
\ee
Since the signature of a spin manifold is a multiple of $16$, the above expression correctly reproduces Witten's result when $\mathcal{M}$ admits a spin structure; otherwise, in general, there is an extra non-trivial $\sigma$-dependent phase in the $S$-duality transformation.\footnote{This phase has also been derived in \cite{Olive:2000yy} via Poisson resummation and in \cite{Etesi:2010sb} by a direct computation of the partition function.}

As is well known, if the space-time $\mathcal{M}$ is a spin manifold, the quantum Maxwell theory has an exact perturbative symmetry that shifts the theta angle by multiples of $2\pi$, which corresponds to the statement $Z[\tau]=Z[\tau+1]$. The latter, together with $S$-duality, makes the partition function into a non-holomorphic modular form of (generally half-integral) weights $(\tfrac{\chi-\sigma}{4},\tfrac{\chi+\sigma}{4})$ under the modular group $SL(2,\mathbb{Z})$. In contrast, in the absence of space-time spin structures, the extra $\sigma$-dependent phase in \eqref{Sduality} prevents the partition function from transforming as a genuine modular form, and, moreover, only theta-angle shifts under multiples of $4\pi$ leave the theory invariant. Therefore, for non-spin $\mathcal{M}$, the duality group is
\be
\langle S,T^2\rangle &=&\left\{M\in SL(2,\mathbb{Z})\;\;{\Big|}\;\;M=\left(\begin{array}{cc}1&0\\ 0&1\end{array}\right)\;{\rm mod}\,2\quad{\rm or}\quad M=\left(\begin{array}{cc}0&1\\ 1&0\end{array}\right)\;{\rm mod}\,2\right\},
\ee
which is the index-three\footnote{This can be easily seen by noticing that its mod-2 reduction (i.e.~its quotient by the subgroup of all matrices congruent modulo 2 to the identity) is $\langle S,T^2\rangle\,{\rm mod}\,2\simeq\mathbb{Z}_2\subset \mathbb{Z}_2\ltimes \mathbb{Z}_3\simeq SL(2,\mathbb{Z}_2)$.} congruence subgroup of $SL(2,\mathbb{Z})$ generated by the order-four element and the free element
\be
S=\left(\begin{array}{cc}0&-1\\ 1&0\end{array}\right)\qquad {\rm and}\qquad T^2=\left(\begin{array}{cc}1&2\\ 0&1\end{array}\right)
\ee
respectively. Under the (standard fractional linear) action of $M=\left(\begin{smallmatrix}a & b\\c & d\end{smallmatrix}\right)\in\langle S,T^2\rangle$ on $\tau$, the partition function transforms as follows
\be\label{DualityNonSpin}
Z[M(\tau)=\tfrac{a\tau+b}{c\tau +d}]&=&e^{\frac{\pi{\rm i}\sigma}{4}s(M)}\,(c\tau+d)^{\frac{\chi-\sigma}{4}}\,(\overline{c\tau+d})^{\frac{\chi+\sigma}{4}}\,Z[\tau]\nonumber\\ \nonumber \\ &=& |c\tau+d|^{\frac{\chi}{2}}\, e^{\frac{{\rm i}\sigma}{4}[\pi s(M)-2\,{\arg}(c\tau+d)]}\,Z[\tau]\,,
\ee
where $s(M)\in\mathbb{N}$ is the number of occurrences of the generator $S$ in the element $M$ (e.g.~$s(-I)=2$, with $I$ the identity matrix). Note that, this quantity is well defined modulo $4$ in the subgroup $\langle S,T^2\rangle\subset SL(2,\mathbb{Z})$, because the latter has a single relation, $S^4=I$. As one can readily check, the transformations \eqref{DualityNonSpin} associated with both $M=\pm I$ are trivial for any choice of $\mathcal{M}$, so that the quantum theory is well defined and invariant under charge conjugation. Moreover, under $s(M)\to s(M)+4$, one must send ${\rm arg}(c\tau+d)\to{\rm arg}(c\tau+d)+2\pi$, so that the phase in Eq.~\eqref{DualityNonSpin} is well defined.

\subsection{Symmetries and anomalies}\label{SymAnom}

\subsubsection{The duality bundle}\label{dualitybundlesec}
The discussion of the previous section makes it clear that the Maxwell partition function $Z[g,\tau]$ is generally not single-valued under continuous deformations of the coupling \cite{Seiberg:2018ntt}. Indeed, electromagnetic duality implies that the space of couplings corresponding to inequivalent Maxwell theories is not $\mathbb{H}$ but the orbifold $\mathbb{H}/\Gamma$, where, as we have seen, $\Gamma$ is either the full $SL(2,\mathbb{Z})$ or a congruence subgroup thereof, when the space-time $\mathcal{M}$ is spin or non-spin respectively. Monodromies of $Z$ are detected by looping around the singularities of $\mathbb{H}/\Gamma$ and, at fixed basepoint $\tau$, they form an Abelian representation of $\Gamma$:
\be\label{ZMonodrom}
Z[M(\tau)]&=& f_\tau(M)\,Z[\tau]\,,
\ee
where $M\in\Gamma$ and $f_\tau(M)\in \mathbb{C}^*$.\footnote{As explained in \cite{Seiberg:2018ntt}, $Z$ should be thought of as a section of the equivariant line bundle defined on $\mathbb{C}\times\mathbb{H}$ by the action $(z,\tau)\to (f_\tau(M)\cdot z,M(\tau))$, with the composition law $f_\tau(M\circ  N)=f_{N(\tau)}(M)\cdot f_\tau(N)$. By choosing local counterterms appropriately, in the spin case, $f_\tau$ can be turned into a $\tau$-independent twelfth root of unity.}

Dualities, in general, are not symmetries, because they relate different theories (or theories corresponding to different values of the parameters). However, in the Maxwell case, as is well known, the duality group $\Gamma$ does not act freely on $\mathbb{H}$. The fixed points\footnote{While $\tau={\rm i},{\rm i}\infty$ are fixed points for both $SL(2,\mathbb{Z})$ and $\langle S,T^2\rangle$, the so-called triality point $\tau=e^{\pi{\rm i}/3}$ is a property of $SL(2,\mathbb{Z})$ only.} are obviously the singularities of $\mathbb{H}/\Gamma$, and duality becomes a global symmetry of the theories associated to these specific values of the coupling. Consequently, non-trivial monodromies of the form \eqref{ZMonodrom}, which, as shown in \eqref{DualityNonSpin}, depend on the gravitational background through $\chi$ and $\sigma$, signal the presence of a mixed 't Hooft anomaly between the duality-symmetry and gravity.\footnote{Since we do not discuss pure 't Hooft anomalies of the duality-symmetry (analyzed in \cite{Hsieh:2019iba,Hsieh:2020jpj}), we only consider constant couplings and trivial $\Gamma$ bundles on space-time.} Moreover, since $Z$ is single-valued on $\mathbb{H}$, they force the partition function to vanish at the fixed points. In what follows, we would like to ask whether the converse holds true, that is, if zeros of $Z$ imply the presence of non-trivial monodromies, and hence of anomalies. Since $Z$ is a \emph{non-holomorphic} section, its vanishing does not guarantee the non-triviality of the equivariant line bundle. As we will see, however, for certain favorable choices of $(\mathcal{M},g)$, for which the conventional duality-symmetries are anomaly free, we will still be able to relate the vanishing of $Z$ to the presence of a mixed anomaly, but for a much subtler symmetry.

Fortunately, the partition function \eqref{PFMaxwell} of Maxwell theory can be computed very explicitly. Following \cite{Witten:1995gf}, one first divides $F$ into its exact and its harmonic piece, and consequently splits the path integral into a continuous part (independent of the theta angle) and a discrete sum over isomorphism classes of line bundles. If we normalize \eqref{PFMaxwell} by the volume of the gauge group and expand the harmonic part of $F$ in a basis $\{e_i\}_i$ of $H^2(\mathcal{M},2\pi \mathbb{Z})$, the computation gives\footnote{Since our main focus is on the $\tau$ dependence of the partition function, we suppress $\tau$-independent factors: They amount to the (metric-dependent) determinant of the laplacian over the non-zero modes of $A$ and to a multiplicative constant given by the volume of a $b_1$-dimensional torus. The latter originates from path integrating over the harmonic part of $A$, consisting of $b_1$ periodic variables, which the theory is completely blind to, due to its (electric) one-form symmetry.}
\be\label{ExplicitPF}
Z[g,\tau]&=&({\rm Im}\,\tau)^{\frac{1}{2}(b_1-1)}\,\sum_{{\bf m}\in \mathbb{Z}^{b_2}}q^{\frac{1}{4}m^i (G_{ij} - Q_{ij})m^j}\bar{q}^{\frac{1}{4}m^i (G_{ij} + Q_{ij})m^j}\,,
\ee
where $b_i$ is the $i$th Betti number of $\mathcal{M}$, $q=e^{2\pi{\rm i}\tau}$, and
\be\label{Q&G}
Q_{ij}=\frac{1}{4\pi^2}\int_{\mathcal{M}}e_i\wedge e_j\,,\qquad G_{ij}=\frac{1}{4\pi^2}\int_{\mathcal{M}}e_i\wedge \star
e_j
\ee
are the intersection form and the induced metric on $H^2(\mathcal{M},\mathbb{Z})$ respectively.  The power of ${\rm Im}\,\tau$ in \eqref{ExplicitPF} arises from the non-zero modes of $A$ modulo gauge transformations,  yielding a gaussian integral over $B_1-b_1-B_0+1$ real variables in the lattice regularization. While mostly irrelevant for our purposes, this factor (in particular its divergence at zero coupling for $b_1>1$) has an amusing physical interpretation, best apparent when our Euclidean space-time $\mathcal{M}$ has a factorized circle.\footnote{We thank Joseph Minahan for discussions on this point.} Indeed, consider Maxwell theory on $\mathcal{N} \times \mathbb{R}$, where $\mathcal{N}$ is an oriented three-manifold with no torsional cycles, and $\mathbb{R}$ plays the role of ``time''. The Hilbert space of (classical) vacua of this theory is the one of a quantum particle propagating on the torus $\mathbb{T}^{b_1(\mathcal{N})}$, which has infinite dimension for $b_1(\mathcal{N})>0$. After compactifying the time, in the $e^2\to 0$ limit, the partition function precisely computes the dimension of this Hilbert space. Since $b_1(\mathcal{M}) = b_1(\mathcal{N}\times S^1) = b_1(\mathcal{N})+1$, this explains the zero-coupling divergence of \eqref{ExplicitPF}.

In the following, we are going to restrict our attention to the discrete sum in \eqref{ExplicitPF} and, in particular, to its two main characters \eqref{Q&G}. As a simplifying technical assumption, we will limit ourselves to considering simply-connected space-time manifolds, i.e.~$b_1=0$, and normalize the partition function by the one corresponding to $\mathcal{M}=\mathcal{S}^4$, which is just $(\sqrt{{\rm Im}\,\tau})^{-1}$. Therefore, we will study the quantity
\be\label{NormalZ}
\tilde{Z}[g,\tau]&=& \sqrt{{\rm Im}\,\tau} \, Z[g,\tau]=\sum_{{\bf m}\in \mathbb{Z}^{b_2}}q^{\frac{1}{4}m^i (G_{ij} - Q_{ij})m^j}\bar{q}^{\frac{1}{4}m^i (G_{ij} + Q_{ij})m^j}\,,
\ee
and its zeros at finite values of the coupling.

First of all, it is easy to see that the quadratic forms $G\pm Q$ appearing above are both positive semi-definite, with kernels intersecting only at the origin, which guarantees the convergence of the sum. However, as pointed out earlier, its non-holomorphic $\tau$ dependence makes it difficult to infer the presence of anomalies from its vanishing. Hence, we find it natural to ask under which circumstances the quantity \eqref{NormalZ} splits into the product of a holomorphic (antiself-dual) and a antiholomorphic (self-dual) part
\be\label{SplitPF}
\tilde{Z}[g,\tau]&=&\sum_{{\bf m_-}\in \mathbb{Z}^{b_2^-}}q^{-\frac{1}{2}m_-^i Q^-_{ij} m_-^j}\,\cdot \sum_{{\bf m_+}\in \mathbb{Z}^{b_2^+}}\bar{q}^{\frac{1}{2}m_+^i Q^+_{ij} m_+^j}\,,
\ee
where we have defined the intersection pairings $Q^\pm=\tfrac12(Q\pm G)$. The two parts are physically interpreted as the instanton and the anti-instanton sum. Such a splitting occurs if two conditions are simultaneously satisfied: 
\begin{enumerate}
	\item $Q$ is block diagonal (over the integers), i.e.~it is of the form $Q=Q^+ \oplus Q^-$, where $Q^+$ is positive definite and $Q^-$ is negative definite;
	\item There exists a metric $g$ on $\mathcal{M}$ whose associated Hodge star is diagonal in the basis (over the reals) where $Q$ is block diagonal, which means that $Q^+,Q^-$ are the spaces of harmonic self-dual and antiself-dual two-forms respectively.
\end{enumerate}

The former condition is a purely topological one, and it amounts to requiring the existence of a matrix $L\in GL(b_2,\mathbb{Z})$ such that $L^TQL=Q^+\oplus Q^-$.\footnote{Such changes of basis are called isomorphisms of intersection forms. Note that, in general, $Q$ and $Q^+\oplus Q^-$ will \emph{not} be similar as endomorphisms.} For example, the intersection form of the product manifold $S^2\times S^2$ is $\left(\begin{smallmatrix}0 & 1\\1 & 0\end{smallmatrix}\right)$, which is clearly not diagonalizable over the integers, whereas the intersection form of the twisted bundle $S^2\tilde{\times} S^2$ is $\left(\begin{smallmatrix}0 & 1\\1 & 1\end{smallmatrix}\right)$, which can be diagonalized to $\left(\begin{smallmatrix}1 & 0\\0 & -1\end{smallmatrix}\right)$ by means of $L=\left(\begin{smallmatrix}0 & -1\\1 & 1\end{smallmatrix}\right)\in SL(2,\mathbb{Z})$. For further details about isomorphisms of intersection forms, see Appendix \ref{MathToolkit}.

As for the latter condition, let us spend a few words on it here, just to build some intuition, and postpone again a more detailed discussion to Appendix \ref{MathToolkit}. Since the metric $G$ is positive definite, we have, in obvious notation, $G({\bf m}\pm\star {\bf m}, {\bf m}\pm\star {\bf m})\geq 0$, with equality holding iff ${\bf m}$ is (anti)self-dual. But from \eqref{Q&G} we realize that $G({\bf m},\star {\bf m})=Q({\bf m},{\bf m})$. Thus we conclude that
\be\label{Bound}
G({\bf m},{\bf m})&\geq & |Q({\bf m},{\bf m})|\,,
\ee
with equality holding precisely (and exclusively) in the space of  (anti)self-dual harmonic two-forms. We dub ``extremal'' the metrics\footnote{More precisely, one should talk about whole conformal classes of metrics, because $G$ is invariant under Weyl rescaling.} that saturate the bound \eqref{Bound}. For example, on the product manifold $S^2\times S^2$, conformal classes of metrics are labeled by a single real number $R$, given by the ratio between the radii of the two spheres. Extremal metrics are those with $R=1$, i.e.~for which the two spheres have equal size. The existence of extremal metrics on generic four-manifolds is an open conjecture in the math literature, proven only recently in specific cases \cite{Katz2003FourmanifoldSA,scaduto2023metricstretchingperiodmap}. The reader interested in learning more about this problem is referred to Appendix \ref{MathToolkit} and references therein.

\subsubsection{A novel duality-symmetry}\label{new symmetry}

Let us then focus on pairs $(\mathcal{M},g)$ that satisfy the two above conditions, and analyze the consequences on the anomaly of duality-symmetries. As we said, in these cases the normalized partition function of Maxwell theory splits as
\be\label{FactorizedZ}
\tilde{Z}[\tau, \bar \tau]= Z_-[\tau] \, Z_+[\bar{ \tau}] \,,
\ee
where the $+(-)$ recalls that we are summing over (anti)self-dual configurations. The two factors, separately, would represent the normalized partition function of Maxwell theory on a space $\mathcal{M}$ equipped with a negative/positive-definite lattice of integral harmonic two-forms. However, due to Donaldson's theorem (see Appendix \ref{MathToolkit}), such a space can never be smooth, unless it is (homeomorphic to) a connected sum of $\overline{\mathbb{CP}}^2/\mathbb{CP}^2$ manifolds (the bar denotes reversed orientation), for which the intersection form is just $-/+$ the identity. Maxwell theory on a connected sum of $N$ copies of $\mathbb{CP}^2$ (or of its orientation reversal) is free of duality anomalies, because its normalized partition function is the $N$th power of the Jacobi theta function
\be
\tilde{Z}_{\mathbb{CP}^2}[\bar{\tau}]&=&\bar{\vartheta}_3(\bar{\tau})\,,
\ee
which vanishes nowhere.\footnote{Note that, if on $\mathbb{CP}^2$ we include the coupling ${\rm i}\pi\int_{\mathcal{M}}c_1\cup w_2$, the normalized partition function becomes $\tilde{Z}_{\mathbb{CP}^2}[\bar{\tau}]=\bar{\vartheta}_4(\bar{\tau})$, which also vanishes nowhere, but it does not simply pick a $\tau$-dependent factor under $S$-duality.} Therefore, in all interesting cases, only the product \eqref{FactorizedZ} will have the physical meaning of a normalized Maxwell partition function, and not each factor separately.

Arguably, the easiest example featuring a partition function which does have a vanishing locus, but admits no non-trivial monodromies under the conventional duality-symmetries is as follows. Start from $\mathcal{M}=(S^2\times S^2)^{\# 8}$, that means the connected sum\footnote{See Appendix \ref{MathToolkit} for all relevant definitions.} of eight copies of $S^2\times S^2$. This is a spin manifold, so the duality group is the entire $SL(2,\mathbb{Z})$. Remarkably, its intersection form can be block diagonalized over the integers, with $Q^\pm=[\pm E_8]$, the Cartan matrix of the $E_8$ algebra, with all entries positive (see Eq.~\eqref{E8matrix}). We call $M_{E_8}\# \overline{M}_{E_8}$ the manifold with intersection form $[+E_8]\oplus [-E_8]$, which is homeomorphic (in fact diffeomorphic) to $(S^2\times S^2)^{\# 8}$, and thus smooth.
Moreover, a metric whose Hodge star is diagonal in this basis is conjectured to exist (see Appendix \ref{MathToolkit} for further details). With this special choice of metric, the normalized partition function of Maxwell theory on $M_{E_8}\# \overline{M}_{E_8}$ reads
\be\label{ZS2S28}
\tilde{Z}_{M_{E_8}\# \overline{M}_{E_8}}[\tau,\bar{\tau}]&=& E_4[\tau] \, \bar{E}_4[\bar{\tau}]  \,,
\ee
where $E_4[\tau]$ is the weight-four Eisenstein series, i.e.~a modular form transforming under $SL(2,\mathbb{Z})$ as $E_4[\tau]\to (c\tau+d)^4 E_4[\tau]$.\footnote{Notice that this is compatible with Eq.~\eqref{DualityNonSpin}, using that $\chi=18$ and $\sigma=0$ for $(S^2\times S^2)^{\# 8}$, and recalling that we have normalized $Z$ as in \eqref{NormalZ}.} Clearly, the particular structure of \eqref{ZS2S28} kills all non-trivial monodromy phases for the ordinary duality-symmetries, and thus naively the theory looks free of anomalies. Nevertheless, the partition function vanishes at the triality point $\tau=e^{\pi {\rm i}/3}$, because $E_4$ does.
This observation poses the problem of how to interpret this zero of the partition function when there is no obvious anomaly argument for it.

It is not difficult to realize that here the equivariant line bundle controlling the anomaly, which is trivial, splits into the product of a non-trivial \emph{holomorphic} line bundle times its antiholomorphic counterpart (which is isomorphic to its dual, or inverse). $E_4[\tau]$ is a holomorphic section of this holomorphic line bundle. We claim that, besides the conventional duality-symmetries at $\tau={\rm i},e^{\pi{\rm i}/3}$, which are both non-anomalous, Maxwell theory on $\mathcal{M}=M_{E_8}\# \overline{M}_{E_8}$ possesses extra duality-symmetries at the same fixed points, which rotate the self-dual fluxes only, leaving the antiself-dual ones untouched (or equivalently, the other way around). These exotic ``partial'' symmetries have the same order of their ordinary counterparts; the one at $\tau={\rm i}$ is still non-anomalous, whereas the one at $\tau=e^{\pi{\rm i}/3}$ has an anomalous phase $e^{4\pi{\rm i}/3}$ under the $\mathbb{Z}_6$ transformation $M=TS=\left(\begin{smallmatrix}1 & -1\\1 & 0\end{smallmatrix}\right)$, associated to the holomorphic line bundle. The anomaly of this exotic triality-symmetry is the reason why the partition function vanishes at the triality point.

To justify our claim, let us focus on the pure $S$ transformation, for simplicity. Recall that, at the level of the normalized partition function $\tilde{Z}$ \eqref{NormalZ}, electromagnetic duality can be rephrased in terms of a Poisson resummation. In particular, for the split normalized partition function \eqref{SplitPF}, we have two separate Poisson resummations:
\be\label{PoissonResum}
\tilde{Z}[\tau,\bar{\tau}]&=&\sum_{{\bf m_-}\in \mathbb{Z}^{b_2^-}}e^{-\pi {\rm i}\tau Q^-({\bf m_-},{\bf m_-})} \sum_{{\bf m_+}\in \mathbb{Z}^{b_2^+}}e^{-\pi {\rm i}\bar{\tau} Q^+({\bf m_+},{\bf m_+})}\nonumber\\ 
&=&(-{\rm i}\tau)^{-\frac{b_2^-}{2}} \sum_{{\bf k_-}\in \mathbb{Z}^{b_2^-}}e^{-\pi {\rm i}(-\frac{1}{\tau}) Q^-({\bf k_-},{\bf k_-})} ({\rm i}\bar{\tau})^{-\frac{b_2^+}{2}} \sum_{{\bf k_+}\in \mathbb{Z}^{b_2^+}}e^{-\pi {\rm i}(-\frac{1}{\bar{\tau}}) Q^+({\bf k_+},{\bf k_+})}\nonumber\\ &=& (-{\rm i}\tau)^{-\frac{b_2^-}{2}} ({\rm i}\bar{\tau})^{-\frac{b_2^+}{2}} \, \tilde{Z}[-\tfrac{1}{\tau},-\tfrac{1}{\bar{\tau}}]\,,
\ee
where we have used the unimodularity of the lattice, that is, ${\rm det}\,Q^+=(-1)^{b_2^-}{\rm det}\,Q^-=1$. When we are at the self-dual point $\tau={\rm i}$, the above equation reduces to an identity, showing indeed that $S$-duality is never anomalous. However, at $\tau={\rm i}$, and only there, we can also just perform one of the two Poisson resummations and reach the same conclusion. As anticipated, this gives us a new symmetry of the theory at the self-dual point. The exact same argument can be carried over for the triality operation (when $\mathcal{M}$ is spin), leading in that case to possible anomalous phases.

As is clear from the discussion above, if we start from a formulation of Maxwell theory at the self-dual point in terms of ``electric'' variables, while the ordinary $S$-symmetry yields the description in terms of ``magnetic'' variables, the partial $S$-symmetry gives us, instead, a novel formulation of the same theory, featuring a mixed magnetic/electric frame, which forbids the presence of a Lagrangian. This ties up nicely with the fact that, as opposed to the conventional duality-symmetries, these exotic ones do \emph{not} originate from dualities on the conformal manifold; they exist solely at the fixed points! Note, also, that they are not purely internal symmetries of Maxwell theory (like ordinary duality-symmetries are), because they only emerge for very specific metrics of the space-time (see Section \ref{MCGroup} for more on this aspect). 

Clearly, this discussion can be generalized to any choice of $(\mathcal{M},g)$ leading to the splitting \eqref{FactorizedZ}, where the (anti)holomorphic part may itself be composed of several factors, and the antiholomorphic line bundle need not necessarily be the inverse of the holomorphic one: Each factor of $\tilde{Z}$ will enjoy its own duality-symmetry. In fact, the appearance of such partial symmetries goes beyond the factorization \eqref{FactorizedZ}: It occurs whenever $Q$ admits, over the integers, a positive (negative) definite subspace made of self-dual (antiself-dual) harmonic two-forms. A paradigmatic example is the K3 manifold, where, with a suitable choice of metric, the normalized Maxwell partition function can be written as
\be\label{K3PF}
\tilde{Z}_{\rm K3}[\tau,\bar{\tau}]&=&E_4[\tau]^2\, \tilde{Z}_{(S^2\times S^2)^{\# 3}}[\tau,\bar{\tau}]\,.
\ee
Here $\tilde{Z}_{(S^2\times S^2)^{\# 3}}$ is the normalized partition function of Maxwell theory on $(S^2\times S^2)^{\# 3}$, which is free of anomalies, whereas the two $E_4$ factors are due to the sublattice $[-E_8]^{\oplus 2}\subset Q$ of antiself-dual harmonic two-forms. The ordinary triality-symmetry in this case has an anomalous phase $e^{8\pi {\rm i}/3}$, which can as well be detected by operating a partial triality transformation that only involves the antiself-dual integral modes.

Let us conclude by stressing the relevance of isolating an (anti)holomorphic piece from the normalized partition function. When this happens, we can claim that the vanishing of such a factorized piece of the partition function at the triality\footnote{Recall that the normalized partition function never vanishes at $\tau={\rm i}$ and $\tau={\rm i}\infty$.} point \emph{implies} the anomaly of a triality-symmetry, possibly of a partial one. This simply follows from the fact that a holomorphic modular form vanishing at $\tau=e^{\pi{\rm i}/3}$ must be proportional to a power of $E_4[\tau]$.

\subsubsection{Physics of extremal metrics}\label{MCGroup}

This section is devoted to analyzing the physical properties of Maxwell theory on manifolds equipped with extremal metrics.

Recall that the mapping class group of a smooth four-manifold $\mathcal{M}$ is defined as $\pi_0\mathrm{Diffeo}(\mathcal{M})$. This group acts on the second cohomology group as $O(Q,\mathbb{Z})$, the subgroup of $GL(b_2,\mathbb{Z})$ that fixes the intersection form $Q$ of $\mathcal{M}$.\footnote{For details on this, the reader is referred to Appendix \ref{MathToolkit}.} Under the action of the mapping class group, the metric $G$ transforms as $G\to L^T G L$. Hence, it is natural to ask what happens if $L\in O(Q,\mathbb{Z})$ leaves $G$ invariant. 

Although in general it is not guaranteed that any matrix $L\in O(Q,\mathbb{Z})$ can be realized as a diffeomorphism (it is nevertheless guaranteed in a large class of four-manifolds), consider one that is, and suppose that it leaves $G$ invariant. This does not guarantee yet that $L$ can be realized as an isometry for any given metric $g$ associated to $G$. It can be shown, however, that the diffeomorphism preserves a Riemannian metric if and only if it is isotopic to a diffeomorphism of finite order.\footnote{See the mathoverflow discussion \href{http://mathoverflow.net/questions/373285}{``Realizing mapping classes as isometries?"}.} Therefore, modulo all these subtleties, we can think of the intersection $O(G,\mathbb{Z})\cap O(Q,\mathbb{Z})$ as an accidental space-time symmetry of the theory. In fact, note that the group $O(G,\mathbb{Z})\cap O(Q,\mathbb{Z})$ is finite because $O(G,\mathbb{Z})$ is finite (being $G$ positive definite). This means, in particular, that all the elements in this intersection have finite order, which, as said, makes them good candidates to be realizable as isometries.

To be more concrete, let us give a couple of examples.
\begin{itemize}
	\item  Consider the product $S^2\times S^2$. As already mentioned, the intersection form is $H=\left(\begin{smallmatrix}0 & 1\\1 & 0\end{smallmatrix}\right)$, and a conformal class of metrics gives $G=\mathrm{diag}(R,1/R)$, where $R$ is the ratio between the sizes of the two spheres. The mapping class group is generated by $H$ itself and minus the identity: $O(H,\mathbb{Z}) = \langle H, -1 \rangle\simeq\mathbb{Z}_2\times\mathbb{Z}_2$. The element $-1$ is not very special: It always leaves any $G$ invariant in any manifold with any intersection form. Indeed, the action of the mapping class group on $G$ is projective. The isometry realizing this element flips the orientation of all the two-cycles of the four-manifold. More interesting, instead, is the element $H$, whose action consists of interchanging the two spheres, sending $R \to 1/R$ \cite{Verlinde:1995mz}. This does not leave $G$ invariant, unless $R=1$, which is exactly the condition that defines the extremal metrics. In this case, the metric $g=g_{S^2}\oplus g_{S^2}$ is manifestly invariant upon interchanging the two spheres (because they have equal radius), and hence this element of the mapping class group corresponds to an extra isometry of the space-time.
	
	\item Consider the connected sum $\mathbb{CP}^2\#\overline{\mathbb{CP}}^2$. The intersection form is given by $Q=\mathrm{diag}(1,-1)$ and the most general metric on the second cohomology is given by $G=\left(\begin{smallmatrix}
		x & \pm \sqrt{x^2 -1} \\
		\pm \sqrt{x^2-1} & x
	\end{smallmatrix}\right)$, where $x\geq 1$, which can be easily derived by imposing $\star^2=1$. The minimum value $x=1$ corresponds to extremal metrics. The mapping class group in this case is given by $O(Q,\mathbb{Z})=O(1,1,\mathbb{Z})=O(1,\mathbb{Z})\times O(1,\mathbb{Z}) = \mathbb{Z}_2\times \mathbb{Z}_2$, which is generated by $-1$ and $P=\mathrm{diag}(1,-1)$. The $-1$ transformation is analogous to the previous example. The generator $P$, instead, which corresponds to flipping the orientation of $\mathbb{CP}^1 \subset\overline{\mathbb{CP}}^2 $, acts on $G$ by interchanging the branches of the square roots. This shows again that the extremal metrics are the only fixed points of the mapping class group. If this transformation is induced by a finite-order diffeomorphism, a theory on $\mathbb{CP}^2\#\overline{\mathbb{CP}}^2$ equipped with an extremal metric enjoys an extended space-time symmetry.  
\end{itemize}
In general, if $Q=Q^+\oplus Q^-$ over the integers, a generic $G$ will not be left invariant under $O(Q,\mathbb{Z})$. However, the $G$ bilinears associated to extremal metrics are left invariant by the subgroup $O(Q^+,\mathbb{Z})\times O(Q^-,\mathbb{Z})$. This is clear from the fact that, by definition, extremal metrics are such that $G=Q^+\oplus -Q^-$. If these transformations can be promoted to isometries, theories on manifolds equipped with extremal metrics experience an enhancement of the space-time symmetry.

Let us now come back to Maxwell theory and draw an analogy between the mapping class group and the duality group.\footnote{See \cite{Verlinde:1995mz}, where these two concepts are elegantly unified in six dimensions by the theory of a self-dual two-form.} The partition function is obviously invariant under diffeomorphisms but, as for the duality group, such transformations are not ``symmetries'', because they relate the theory on $(\mathcal{M},g)$ to the theory on $(\mathcal{M},g'=\phi(g))$ with $\phi\in{\rm Diff}(\mathcal{M})$. This clearly holds for discrete diffeomorphisms too, i.e.~for those induced by elements of the mapping class group: Suppose a given $L\in O(Q,\mathbb{Z})$ is realized by a diffeomorphism $\phi_L\in {\rm Diff}(\mathcal{M})$, then
\be
Z[\phi_L(g),\tau]=Z[g,\tau]\,.
\ee
Now, pick an extremal metric $g_*$. This will induce a bilinear $G_*$ with the property that $G_* = L^T G_* L$. As noted above, this feature is likely to promote $\phi_L$ to an isometry of $g_*$, and thus to lead to a (non-anomalous) symmetry of the whole theory (and not just to an invariance of the discrete sum over line bundles). In other words, the extremal metrics $g_*$ play the exact same role for the mapping class group as the orbifold points $\tau_*={\rm i},e^{\pi {\rm i}/3}$ do for the duality group.

The above discussion sheds some light on the very special nature of the exotic duality-symmetries that we introduced in the previous section. They are neither purely internal nor purely external symmetries of Maxwell theory; they are somewhat a mixture of the two, because they emerge exclusively at the fixed points both of the mapping class group (i.e.~the extremal metrics $g_*$) and of the duality group (i.e.~the orbifold points $\tau_*$). Perturbing away from either one of these values, even mildly, breaks this peculiar type of symmetry.

\subsection{Action on lines}\label{ActLines}

In this section, we analyze how extended operators behave under the partial symmetry we have defined in the previous section. We begin by reviewing the implications of the standard $S$-duality for the correlation functions of line operators \cite{Deligne:1999qp},\footnote{See also \cite{Zucchini:2002te}, where the duality covariance of the Wilson loop is analyzed in detail.} and then extend these results to the case of the partial version of the $S$-duality-symmetry.

Let us consider the partition function of Maxwell theory at coupling $\tau$ with the insertion of a general number of Wilson loops. This can be written by inserting a source for the connection as
\be\label{WLco}
Z[g,\tau,j]&=&({\rm Im}\,\tau)^{\frac12(B_1-B_0)}\int \mathcal{D}A \,\exp\left(-S(\tau,F)+{\frac{\mathrm{i}}{2\pi}\!\!\int_\mathcal{M}\!A\wedge j}\right)\,,
\ee
with $j=\sum_k2\pi q_k \delta^{(3)}(\gamma_k)$, where $\delta^{(3)}(\gamma_k)$ is the three-form “delta-function" localized on the one-cycle $\gamma_k$.
With this choice for the source $j$, the partition function with insertion is by definition
\be
Z[g,\tau,j]&=&Z[g,\tau]\,\vev{\prod_k W^{q_k}[\gamma_k]}_\tau\,.
\ee
Considering Wilson loops supported on homologically trivial cycles,\footnote{Correlators of Wilson ('t Hooft) loops supported on non-contractible cycles vanish on a compact manifold, due to the electric (magnetic) one-form symmetry \cite{Gaiotto:2014kfa}.} we have that the source is exact and can be written as $j=\mathrm dJ$. Therefore, integrating by parts we have
\be\label{WLcorr}
Z[\tau,j]&=&({\rm Im}\,\tau)^{\tfrac12(B_1-B_0)}\int \mathcal{D}A \,\exp\left(-S(\tau,F)+{\frac{\mathrm{i}}{2\pi}\!\!\int_\mathcal{M}\! F\wedge J}\right)\,.
\ee
We can now perform $S$-duality on the partition function with sources, using the procedure explained in Section \ref{wittenWork}.
We consider the action \eqref{ExtMaxAct} with the addition of the external-source deformation
\be
\mathcal{S}&=&\frac{{\rm i}}{2\pi}\int_{\mathcal{M}}\tilde{F}\wedge (\mathcal{F}-{\rm d}A)+S(\tau,\mathcal{F})-\frac{\mathrm{i}}{2\pi}\!\!\int_\mathcal{M}\! \mathcal{F}\wedge J\,.
\ee
Again, integrating out $\tilde{F}$ freezes $\mathcal{F}$ to $\mathcal{F}\stackrel{{\rm loc}}{=}{\rm d}A$, and we get back the partition function with sources in the electric frame. The magnetic-dual frame is obtained by integrating out $A$, so that $\tilde{F}$ localizes on closed two-forms with integral periods. In this way we obtain \eqref{ExtZ} with the insertion of the source, namely the action \eqref{DualMaxAct} is modified to
\be
\mathcal{\tilde{S}}&=&\frac{{\rm i}}{4\pi}\int_{\mathcal{M}} \bar{\tau}\,\mathcal{F}^+\wedge \mathcal{F}^++\tau\, \mathcal{F}^-\wedge\mathcal{F}^-+2\,(\tilde{F}^+- J^+)\wedge \mathcal{F}^++2\,(\tilde{F}^--J^-)\wedge \mathcal{F}^-\nonumber\\ \nonumber\\ &=& S(1,\mathcal{\hat G})+S(-\tfrac{1}{\tau},\hat{F})\,
\ee
where we have defined
\be
\hat{F}=\tilde{F}- J\,,\qquad\mathcal{\hat G}^+=\sqrt{\bar\tau} \,\mathcal{F}^++\frac{1}{\sqrt{\bar\tau}}\,\hat{F}^+\,,\qquad \mathcal{\hat G}^-=\sqrt{\tau} \,\mathcal{F}^-+\frac{1}{\sqrt{\tau}}\,\hat{F}^-\,.
\ee
After the integration over $\mathcal{\hat G}$, taking into account the $\tau$ dependence we get

\be\label{disorder insertion}
Z[\tau,j]=({\rm i}\bar\tau)^{-\frac{B_2^+}{2}}\,(-{\rm i}\tau)^{-\frac{B_2^-}{2}}({\rm Im}\,\tau)^{\frac12(B_1-B_0)}\int \mathcal{D}\hat{A} \,e^{-S(-\frac{1}{\tau},\hat{F})}\,,
\ee
where the integration variable $\hat{A}$ is a connection with Dirac-string singularities. More explicitly we have that
\be
\mathrm d\hat{F}=\mathrm d\tilde{F}- \mathrm dJ=-2\pi\sum_kq_k\delta^{(3)}(\gamma_k)\,,
\ee
where in the second equation we used that $\tilde{F}$ is closed due to the path integration over $A$. This is exactly the disorder-operator description of the insertion in the functional integral of 't Hooft lines of charge $-q_k$ along the cycles $\gamma_k$.
Therefore, using \eqref{WLcorr}, we can rewrite \eqref{disorder insertion} as
\be
Z[\tau]\,\vev{\prod_k W^{q_k}[\gamma_k]}_\tau=e^{-\frac{\pi{\rm i}\sigma}{4}}\,(\tau)^{-\frac{\chi-\sigma}{4}}\,(\bar\tau)^{-\frac{\chi+\sigma}{4}}\,Z[-\tfrac1\tau]\vev{\prod_k T^{-q_k}[\gamma_k]}_{-\frac{1}{\tau}}\,.
\ee
Assuming $Z[\tau]\neq0$, we can divide both sides by it and, applying \eqref{Sduality}, arrive at the following relation:

\be\label{SdualityLines}
\vev{\prod_k W^{q_k}[\gamma_k]}_\tau=\vev{\prod_k T^{-q_k}[\gamma_k]}_{-\frac{1}{\tau}}\,,
\ee
which, at the self-dual point $\tau=\mathrm i$, implies a selection rule that connects the vevs of Wilson and 't Hooft lines within the same theory. Note that the relation \eqref{SdualityLines} does not see the additional factors of $\tau$ coming from the modular properties of the partition function.

We now generalize this procedure to the partial symmetry described in Section \ref{new symmetry}.
To do so, we start from the expression in \eqref{WLcorr} and, as in Section \ref{dualitybundlesec}, we expand the harmonic part of $F$ in a basis of $H^2(\mathcal{M},2\pi\mathbb{Z})$. In particular, the source deformation is

\be
\frac{\mathrm{i}}{2\pi}\!\int_\mathcal{M}\!\!F\wedge J=\frac{\mathrm{i}}{2\pi}\!\int_\mathcal{M}\!(\mathrm{d}a+m^j e_j)\wedge J= \frac{\mathrm{i}}{2\pi}\!\int_\mathcal{M}\!\mathrm d a\wedge J+2\pi\mathrm{i}m^iQ_{ij} \alpha^j\,,
\ee
where we defined $\displaystyle Q_{ij}\alpha^j= \tfrac{1}{4\pi^2}\int_\mathcal{M}\!e_i\wedge J\in\mathbb R /\mathbb{Z}$.\footnote{The periodicity of the coefficients $\alpha^j$ comes from the observation that $\exp(2\pi{\rm i} m^iQ_{ij} n^j)=1$ when $n^j \in \mathbb{Z}$.} With the same assumptions of Section \ref{dualitybundlesec}, we obtain
\be\label{Disregard}
Z[\tau,j]&=&({\rm Im}\,\tau)^{\frac12(B_1-B_0)}\int \mathcal{D}a\, e^{-\frac{{\rm Im}\,\tau}{4\pi}\int da\wedge \star da+\frac{\mathrm{i}}{2\pi}\int da\wedge J}\cdot\nonumber\\ &&\cdot\sum_{{\bf m}\in \mathbb{Z}^{b_2}}q^{\frac{1}{4}m^i (G_{ij} - Q_{ij})m^j}\bar{q}^{\frac{1}{4}m^i (G_{ij} + Q_{ij})m^j} e^{2\pi\mathrm{i}m^iQ_{ij} \alpha^j}\,.
\ee 

Let us disregard for the moment the continuous part (the first line in the above formula), on which the partial duality acts trivially, and focus instead on the discrete sum. Assuming that the conditions of Section \ref{dualitybundlesec} on $G$ and $Q$ are satisfied, we obtain the split form
\be
\label{partsum}
\tilde{Z}[\tau,J]=\left(\sum_{{\bf m_-}\in \mathbb{Z}^{b_2^-}}e^{-\pi\mathrm{i}\tau (m_-^i Q^-_{ij} m_-^j)+2\pi\mathrm{i}m^i_- Q^-_{ij} \alpha^j_-}\,\right)\left( \sum_{{\bf m_+}\in \mathbb{Z}^{b_2^+}}e^{-\pi \mathrm{i}\bar{\tau}(m_+^i Q^+_{ij} m_+^j) +2\pi\mathrm{i}m^i_+ Q^+_{ij}\alpha^j_+}\right)\,.
\ee
By performing a Poisson resummation, like in Eq.~\eqref{PoissonResum}, but on just one of the factors, for example the first one, we get
\be\label{Oop}
\tilde{Z}[\tau,J]=(-{\rm i}\tau)^{-\frac{b_2^-}{2}}\hspace{-.18cm}\left( \sum_{{\bf k_-}\in \mathbb{Z}^{b_2^-}}e^{-\pi {\rm i}(-\frac{1}{\tau})[(k_--\alpha_-)^i Q^-_{ij} (k_--\alpha_-)^j]}\right)\hspace{-.2cm} \left( \sum_{{\bf m_+}\in \mathbb{Z}^{b_2^+}}e^{-\pi \mathrm{i}\bar{\tau}(m_+^i Q^+_{ij} m_+^j) +2\pi\mathrm{i}\alpha^i_+ Q^+_{ij}m^j_+}\right)\!.
\ee
At the self-dual point $\tau=\mathrm{i}$ the anomalous phase is trivialized, and $-\tfrac{1}{\tau} = \tau$. The structure of \eqref{partsum} is exactly the one of generalized Jacobi theta functions with characteristics \cite{Mumford,Alvarez-Gaume:1986rcs}, and Eq.~\eqref{Oop} directly follows from their modular properties. We will have more to say about these functions in Section \ref{PartialGauging}.

Let us now comment on this result. In general, the insertion of a Wilson or 't Hooft line can be interpreted as turning on a non-flat magnetic or electric background, respectively, for the associated one-form symmetry, given by a Dirac delta localizing on a disk $D$ bounding the line $\gamma$. Then, $\alpha^i\in\mathbb{R}/\mathbb{Z}$ are the components of the harmonic part of this Dirac delta in the basis $\{e_i\}_i$. In other words, for the Wilson line, we have
\be
\int_\gamma A\equiv\int_{D}F=\int_\mathcal{M}  F\wedge\delta^{(2)}(D) =\int_\mathcal{M} {\rm d}a \wedge\star{\rm d}b +\int_\mathcal{M} m^ie_i\wedge \alpha^je_j\,,
\ee
where ${\rm d}\star{\rm d}b=-\delta^{(3)}(\gamma)$,
and the corresponding background for the magnetic one-form symmetry, modulo gauge transformations, reads
\be\label{originalMagnetic}
B_{\rm m}&=&\alpha^i\,e_i+\star{\rm d}b\,.
\ee
It is clear that, the discrete sum in \eqref{Disregard} is associated with a flat magnetic background. Indeed, when neglecting the continuous part, the effect of the Wilson line is equivalent to the insertion of the following magnetic one-form symmetry operator
\be
U_{\rm m}(\alpha)&=&e^{{\rm i}\alpha^i\int_{\Sigma_i}F}\,,
\ee
where $\Sigma_i={\rm PD}(e_i)$, with `PD' denoting Poincar\'e duality.
In this case, the action of the partial symmetry simply consists in exchanging the antiself-dual part of the magnetic background with an electric one, leaving the self-dual part of it untouched. 

In order to see the action of the partial symmetry on the whole Wilson line, we have to reincorporate the continuous part in \eqref{Disregard}, which, however, is left untouched by the symmetry operation. Therefore, what we get after the partial transformation cannot be cast in terms of an insertion of any dyonic line. Nevertheless, we can still express the result in terms of a \emph{restricted} mixed electric-magnetic background of the form
\be
\tilde{B}_{\rm e}&=&\alpha_-^ie^-_i\,,\nonumber\\
\tilde{B}_{\rm m}&=&\alpha_+^ie^+_i+\star{\rm d}b\,,
\ee
where the restriction comes from the allowed gauge transformations: $\tilde{B}_{\rm e}\to \tilde{B}_{\rm e}+h_-$ and $\tilde{B}_{\rm m}\to \tilde{B}_{\rm m}+h_++{\rm d}c$, with $h_\pm$ harmonic (anti)self-dual integral two-forms and $c$ a globally well-defined one-form. Note that $\{h_\pm,c\}$ correspond exactly to the gauge redundancy of the original magnetic background \eqref{originalMagnetic}, i.e.~$B_{\rm m}=\delta^{(2)}(D)$.

\section{The non-invertible case}\label{Sec:Non-Inv}

In Section \ref{Sec:Inv} we have learned a lesson: If the partition function vanishes at the triality point, this is the signal of an anomalous symmetry. In particular, we have proved this statement when the zero originates from an (anti)holomorphic factor. This begs the question: Can the same conclusion be drawn if the partition function vanishes at a different point (i.e.~a smooth one) of the conformal manifold $\mathbb{H}/\Gamma$? The purpose of this section is to build evidence in this direction.

Recall that, the zeros of a (anti)holomorphic modular form of weight $k$ in the fundamental region $\mathbb{H}/SL(2,\mathbb{Z})$ are governed by \cite{lang2001introduction}
\be\label{ZeroHolMod}
\frac{1}{2}\nu_{\rm i} + \frac{1}{3}\nu_\rho + \nu_{{\rm i} \infty}  + \sum_{p\neq {\rm i},\rho,{\rm i}\infty} \nu_p &=& \frac{k}{12} \,,
\ee
where $\nu_x$ is the order of the zero at the point $x$, and $\rho = e^{\frac{{\rm i}\pi}{3}}$ is the triality point. Notice, in particular, that generalized theta functions associated with unimodular lattices have zeros at neither $\tau = {\rm i}$ nor $\tau = {\rm i} \infty$. Moreover, since the polynomial ring of holomorphic modular forms is $\mathbb{C}[E_4, E_6]$, powers of $E_4$ are the only ones that vanish at the triality point. Therefore, if we consider a lattice that does not have any $E_8$ factor (like the one we will encounter in Section \ref{Sec:Leech}), Equation \eqref{ZeroHolMod} gives us information about the zeros of its generalized theta function at smooth points. 

The physical interpretation of these zeros is a bit elusive. One possibility is that they might be linked to anomalies of the non-invertible generalization of the duality-symmetries. Recent studies have indeed shown that, by combining the ordinary $SL(2,\mathbb{Z})$ duality transformations with the gauging of discrete subgroups of the $U(1)\times U(1)$ one-form symmetry, one gets non-invertible duality operations that enhance the action on $\tau$ to $SL(2,\mathbb{Q})$ \cite{Niro:2022ctq,Sela:2024okz}.\footnote{The enhancement is even to the full $SL(2,\mathbb{R})$, as shown in \cite{Hasan:2024aow,Paznokas:2025epc}.} Consequently, besides the ordinary invertible duality-symmetries $\mathbb{Z}_2$ and $\mathbb{Z}_3$ at the orbifold points, we will have non-invertible duality-symmetries associated to the $PSO(2,\mathbb{Q})$ (or even to the $PSO(2,\mathbb{R})$) stabilizer group of smooth points.

After reviewing the construction of non-invertible dualities in Section \ref{ReviewGauging}, we will discuss how to implement ``partial'' gauging operations in Section \ref{PartialGauging}, which allow us to turn non-invertible the partial symmetries that we have introduced. In Section \ref{Sec:Leech}, we will present an explicit example of partition function vanishing at a smooth point, and interpret it from the anomaly viewpoint.

\subsection{Non-invertible dualities}\label{ReviewGauging}

In this section, we make the first step in the construction of the partial non-invertible symmetry of Maxwell theory by reviewing the results of \cite{Choi:2021kmx,Kaidi:2021xfk,Choi:2022zal,Cordova:2023ent} and \cite{Niro:2022ctq,Sela:2024okz}. In particular, we are going to review the action of the discrete gauging on the coupling and the construction of the non-invertible symmetry operator.

First, let us discuss the discrete gauging. If we turn on backgrounds for both the electric and magnetic one-form symmetries, and sum over them, the partition function becomes\footnote{For the normalization we use the same conventions of \cite{Gaiotto:2014kfa}.}
\be\label{eq:discretebackground}
Z_{G_{\frac{N_{\rm e}}{N_{\rm m}}}}[\tau]&=&
\frac{|H^0(\mathcal{M}, \mathbb{Z}_{N_{\rm e}})|}{|H^1(\mathcal{M}, \mathbb{Z}_{N_{\rm e}})|}\frac{|H^0(\mathcal{M},\mathbb{Z}_{N_{\rm m}})|}{|H^1(\mathcal{M},\mathbb{Z}_{N_{\rm m}})|}\sum_{B_{\rm m}\in H^2(\mathcal{M},\mathbb{Z}_{N_{\rm m}})} \sum_{B_{\rm e}\in H^2(\mathcal{M}, \mathbb{Z}_{N_{\rm e}})} (\mathrm{Im}\, \tau )^{\frac{1}{2}(B_1 - B_0)}\nonumber \\ &\cdot& \int \mathcal{D}A\,
e^{-\frac{1}{2e^2}\int_\mathcal{M}( F -B_{\rm e} )\wedge \star (F-B_{\rm e}) -\frac{{\rm i}\theta}{8\pi^2}\int_\mathcal{M} (F-B_{\rm e})\wedge (F-B_{\rm e}) +\frac{\rm i}{2\pi}\int_\mathcal{M}(F-B_{\rm e})\wedge B_{\rm m}} \ ,
\ee
where we dubbed $G_{\frac{N_{\rm e}}{N_{\rm m}}}$ the gauging operator, following the convention in \cite{Hasan:2024aow}. In Eq.~\eqref{eq:discretebackground}, $B_{\rm e}$ is the background of the electric $\mathbb{Z}_{N_{\rm e}}$ and $B_{\rm m}$ is the background of the magnetic $\mathbb{Z}_{N_{\rm m}}$ one-form symmetries. When $\gcd(N_{\rm e}, N_{\rm m})=1$, there is no mixed anomaly between the two one-form symmetries, and the sum over the background fields can be carried out safely. This process has the effect of changing the periodicity of the field strength and of its dual \cite{Hayashi:2022fkw,Cordova:2023ent,Hasan:2024aow}. One can go back to a canonically normalized gauge field by simply rescaling the coupling, which leads to the following relation
\be\label{gauged_maxwell}
Z_{G_{\frac{N_{\rm e}}{N_{\rm m}}}}[\tau]& =& N_{\rm m}^\chi \, Z\Big[\frac{N_{\rm m}^2}{N_{\rm e}^2}\,\tau\Big] \ .
\ee
As discussed in the references, the gauging operator is invertible up to condensates, meaning that it enjoys a non-invertible fusion rule
\be\label{FusionRule}
G_{q/n}\, G_{p/m} &=& G_{\frac{pq}{mn}} \, C_{\binom{\gcd{(q,m)}}{\gcd{(p,n)}}} \, ,
\ee
where $C_{\binom{a}{b}}$ is a condensate of electric and magnetic symmetry operators, supported on an interface $\mathcal{M}_3$
\be\label{Condensate}
C_{\binom{a}{b}}& =& \sum_{\begin{smallmatrix} \mathcal{M}_2 \in H_2(\mathcal{M}_3, \mathbb{Z}_{b}) \\ \mathcal{N}_2 \in H_2(\mathcal{M}_3, \mathbb{Z}_{a})\end{smallmatrix}}\exp \Big{(} {\rm i}\,a \oint_{\mathcal{M}_2}  F + {\rm i}\,b \oint_{\mathcal{N}_2} \tilde{F} \Big{)} \, ,
\ee
where $\tilde{F}=\tfrac{2\pi {\rm i}}{e^2}\star F-\tfrac{\theta}{4\pi^2}F$ is the flux measuring the electric charge. The condensate \eqref{Condensate} defines a projector of line operators: It projects out all the Wilson lines whose charge is not a multiple of $a$ and all the 't Hooft lines whose charge is not a multiple of $b$. The gauging operation acts on the coupling constant via the matrix $G_{\frac{N_{\rm e}}{N_{\rm m}}} = \left(\begin{smallmatrix} \frac{N_{\rm m}}{N_{\rm e}} & 0 \\ 0 & \frac{N_{\rm e}}{N_{\rm m}}  \end{smallmatrix}\right)$, effectively extending the $PSL(2,\mathbb{Z})$ action to $PSL(2,\mathbb{Q})$.

Putting all these ingredients together, we have that almost any point in the moduli space (or literaly any point, if we use the $PSL(2,\mathbb{R})$ version) has a non-trivial isotropy group, which is an intrinsic property of the point itself. We can concretely visualize this group by taking any path in the moduli space. For instance, consider a coupling of the form $\tau = {\rm i} \frac{N_{\rm e}^2}{N_{\rm m}^2}$. This value is fixed by an entire $U(1)\cap PSL(2,\mathbb{Q})$ group. We can identify e.g.~a $\mathbb{Z}_2$ subgroup of it by applying the sequence of transformations $D\tau = G_{\frac{N_{\rm m}}{N_{\rm e}}}SG_{\frac{N_{\rm e}}{N_{\rm m}}} \tau = \tau$. In other words, we rescale the coupling via a discrete gauging, perform the standard $S$ duality-symmetry at the self-dual point, and then go back to the original coupling with the opposite discrete gauging. At the level of Maxwell theory, this operation is always well defined, since $Z[{\rm i}]$ is always non-zero. The price we pay for including the discrete gauging is that the operator $D$ now defines a \emph{non-invertible} symmetry of the theory at that specific coupling, because of the fusion rule \eqref{FusionRule} of the $G$ operators.

Of course we can do something analogous to define a $\mathbb{Z}_3$ subgroup of the isotropy group of $\tau = \rho \frac{N_{\rm e}^2}{N_{\rm m}^2}$, by making use of the triality symmetry. However, we must be careful, because we can only do that if the partition function does not vanish at the triality point $\rho$. This vanishing represents an obstruction to constructing the non-invertible triality symmetry operator associated to that specific $\mathbb{Z}_3$ subgroup.
Let us give an example. Consider the K3 manifold. Its partition function \eqref{K3PF} vanishes at the triality point, as a consequence of a mixed anomaly of the triality-symmetry with gravity. 
Starting from a generic coupling $\tau = q^2 e^{\pi {\rm i} /3}$ with $q \in \mathbb{Q}$, we can apply the gauging operator $G_q$, which leads to
\begin{equation}
	Z_{G_{q}}[ q^2 e^{\pi {\rm i} /3}] \sim Z[e^{\pi {\rm i} /3}] = 0 \, ,
\end{equation}
which prevents us from constructing the non-invertible triality operator.

\subsection{``Partial'' gaugings}\label{PartialGauging}

In order to better understand how partial symmetries arise, as well as their possible generalization to partial non-invertible symmetries, it is useful to describe the properties of the field theory in a more algebraic manner. In particular, we show how the partition function of Maxwell theory on a manifold obtained via connected sums can be rewritten as the product of the partition function of Maxwell theory on the individual components, suitably normalized. This allows us, at least formally, to define the partition function even on non-smooth spaces.

To see this, let us recall that a connected sum is achieved by cutting from two closed four-manifolds an $S^3$ and then gluing the two manifolds along this boundary (see Appendix \ref{MathToolkit}). As we discussed in Section \ref{Sec:Inv}, when the intersection matrix of a manifold is block diagonal, one can think of the associated four-manifold as obtained via a connected sum of ``irreducible'' components. Moreover, if the Hodge-star operator is simultaneously block diagonal, the partition function can be written as a product
\be
Z_{\mathcal{M} \# \mathcal{N} }[\tau,\bar{\tau}] &=& ({\rm Im}\,\tau)^{-\frac{1}{2}} \, \tilde{Z}_{\mathcal{M}}[\tau,\bar{\tau}] \tilde{Z}_{\mathcal{N}}[\tau,\bar{\tau}] \, ,
\ee
where $\tilde{Z}_{\mathcal{M}}[\tau,\bar{\tau}]$ and $\tilde{Z}_{\mathcal{N}}[\tau,\bar{\tau}]$ are as in Eq.~\eqref{NormalZ}, for the two manifolds $\mathcal{M}$ and $\mathcal{N}$, respectively, which we assume to be simply-connected. We can now notice that, since the partition function on $\mathcal{S}^4$ is $Z_{\mathcal{S}^4}[\tau,\bar{\tau}]=({\rm Im}\,\tau)^{-\frac{1}{2}}$, by multiplying and dividing by this quantity we end up with\footnote{Recall that, since our focus is on the $\tau$ dependence, we are neglecting the laplacian determinants. The latter would spoil the factorization \eqref{SplitPFS4}, because the laplacian determinant on $\mathcal{M} \# \mathcal{N}$ is in general different from the product of the respective laplacian determinants on $\mathcal{M}$ and $\mathcal{N}$.}
\be\label{SplitPFS4}
Z_{\mathcal{M} \# \mathcal{N} }[\tau,\bar{\tau}] &=& \frac{Z_{\mathcal{M}}[\tau,\bar{\tau}]Z_{\mathcal{N}}[\tau,\bar{\tau}]}{Z_{\mathcal{S}^4}[\tau,\bar{\tau}]}  \, .
\ee
This decomposition is similar to the one in \cite{Witten:1988hf}, with the crucial difference that now the above equation holds for a non-topological theory. The expression \eqref{SplitPFS4} allows us to interpret the partial symmetry in terms of the symmetry of the ``partial'' theory.

As we anticipated, this may become a way to formally define the partition function even on non-smooth spaces, like for example\footnote{From now on, we do not indicate the $\bar{\tau}$ dependence in order not to clutter the notation.}
\be
Z_{M_{E_8} \# \overline{M}_{E_8} }[\tau]& =& \frac{Z_{M_{E_8}}[\tau]Z_{\overline{M}_{E_8} }[\tau]}{Z_{\mathcal{S}^4}[\tau]}  \, .
\ee
Despite this is a mere algebraic manipulation, as we will show shortly, it will allow us to define a notion of gauging of the partial theories, and therefore to construct non-invertible partial symmetries.

To this end, we will follow \cite{Verlinde:1995mz},\footnote{The same expression is obtained in \cite{Alvarez-Gaume:1986rcs} when considering the one-loop partition function of fermions on a genus-$g$ Riemann surface. In that case, $\Omega = {\rm i}\tfrac{2\pi}{e^2} \, G - \tfrac{\theta}{2\pi} \, Q$ is the period matrix of the surface, while $\psi$ and $\phi$ represent different spin structures.} and define the generalized partition function
\be
Z \genfrac[]{0 pt}{0}{\psi}{\phi} &=& 
({\rm Im}\,\tau)^{-\frac{1}{2}}\,\sum_{{\bf m}\in \mathbb{Z}^{b_2}}q^{\frac{1}{4}(m+\psi)^i (G_{ij} - Q_{ij})(m+\psi)^j}\bar{q}^{\frac{1}{4}(m+\psi)^i (G_{ij} + Q_{ij})(m+\psi)^j} e^{2 \pi i (m+\psi)^iQ_{ij}\phi^j} ,
\ee
where $\psi$ and $\phi$ are vectors with rational entries, that correspond to non-trivial backgrounds for finite subgroups of the $U(1)$ electric and magnetic one-form symmetries respectively.\footnote{Turning on $\psi ,  \phi$ is equivalent to turning on background two-form gauge fields $B_{\rm e} , B_{\rm m}$ of the electric, magnetic one-form symmetries. This has the consequence of changing the quantization condition of $F$, from integral to fractional flux. When $B_{\rm e/m} = w_2$, the second Stiefel-Whitney class of the manifold, it changes the statistics of the line operators \cite{Olive:2000yy,Thorngren:2014pza,Kan:2024fuu}.} Gauging is then achieved by summing over these backgrounds.

As an example, let us consider the gauging of a $\mathbb{Z}_2$ subgroup of the electric one-form symmetry for Maxwell theory on $\overline{\mathbb{CP}}^2$. The partition function with fluxes is given by
\be
Z \genfrac[]{0 pt}{0}{{\frac{n}{2}}}{0}&=& ({\rm Im}\,\tau)^{-\frac{1}{2}} \sum_{ m\in \mathbb{Z}}e^{\pi {\rm i} \tau(m+ \frac{n}{2})^2} \,,
\ee
for $n = 0,1$. Summing over all possible backgrounds amounts to 
\be
\sum_{n=0,1} Z \genfrac[]{0 pt}{0}{\frac{n}{2}}{0} [\tau] &=& ({\rm Im}\,\tau)^{-\frac{1}{2}} \Big{(} \sum_{ m\in \mathbb{Z}}e^{\pi {\rm i} \tau m^2} + \sum_{ m\in \mathbb{Z}}e^{\pi {\rm i} \tau(m+ \frac{1}{2})^2} \Big{)}\, \nonumber \\
&=& ({\rm Im}\,\tau)^{-\frac{1}{2}} \Big{(} \sum_{ m\in \mathbb{Z}}e^{\pi {\rm i} \frac{\tau}{4}(2m)^2} + \sum_{m\in \mathbb{Z}}e^{\pi {\rm i} \frac{\tau}{4} (2 m+ 1)^2} \Big{)} = \frac{1}{2} Z \genfrac[]{0 pt}{0}{0}{0} [\tfrac{\tau}{4}]\,,
\ee
which is compatible with \eqref{gauged_maxwell}, using the normalization in \eqref{eq:discretebackground}.

Repeating the same analysis for a generic manifold and an electric $\mathbb{Z}_{N_{\rm e}}$ gauging, we get
\be
Z_{G_{N_{\rm e}}}[\tau] =  N_{\rm e}^{1-b_1} \sum_{{\bf n}} Z \genfrac[]{0 pt}{0}{{\frac{{\bf n}}{N_{\rm e}}}}{0}[\tau] = Z[\tfrac{\tau}{N_{\rm e}^2}] \, ,
\ee
where now ${\bf n}$ is a $b_2$-vector with  $\mathbb{Z}_{N_{\rm e}}$-valued entries, encoding the electric background for each two-cycle in the manifold.

We can repeat the above considerations for the magnetic gauging. Again, as an example, we consider $\overline{\mathbb{CP}}^2$ and turn on a $\phi$-type background
\be
Z \genfrac[]{0 pt}{0}{0}{\frac{n}{2}} &=& ({\rm Im}\,\tau)^{-\frac{1}{2}} \sum_{ m\in \mathbb{Z}}e^{\pi {\rm i} \tau m^2} e^{\pi {\rm i}  n  m} \,,
\ee
with again $n=0,1$. Summing over the two options, we obtain
\begin{align}
	\sum_{n=0,1} Z \genfrac[]{0 pt}{0}{0}{\frac{n}{2}}[\tau] = ({\rm Im}\,\tau)^{-\frac{1}{2}} \sum_{ m\in \mathbb{Z}} (1 + (-1)^{m}) e^{\pi {\rm i} \tau m^2}\, = 2 ({\rm Im}\,\tau)^{-\frac{1}{2}} \sum_{ m\in \mathbb{Z}} e^{\pi {\rm i} \tau (2m)^2}\, = 4 Z \genfrac[]{0 pt}{0}{0}{0}(4 \tau) \, ,
\end{align}
which is compatible with \eqref{gauged_maxwell}, using the normalization in \eqref{eq:discretebackground}. 

For a generic manifold and a magnetic $\mathbb{Z}_{N_{\rm m}}$ gauging, we get
\begin{align}
	Z_{G_{1/N_{\rm m}}}[\tau] = N_{\rm m}^{1-b_1} \sum_{\bf n} Z \genfrac[]{0 pt}{0}{0}{\frac{\bf n}{N_{\rm m}}}[\tau] = N_{\rm m}^{\chi} Z[N_{\rm m}^2\tau] \, ,
\end{align}
where ${\bf n}$ is a $b_2$-vector with  $\mathbb{Z}_{N_{\rm m}}$-valued entries, encoding the magnetic background for each two-cycle in the manifold.

Note that when both $\psi$ and $\phi$ are on, the partition function gets a factor $\exp(\psi Q \phi)$, which might lead to inconsistencies when $\gcd(N_{\rm e},N_{\rm m}) \neq 1$. When the order of the two gauged symmetries are co-prime, one can redefine the backgrounds as follows
\begin{align}
	\psi \to N_{\rm m} \psi  \, ,\qquad \phi \to N_{\rm e} \phi \, .
\end{align}
From the field-theory point of view, this amount to taking the gauge fields of the electric and magnetic symmetries to be of the form $B_{\rm e}=\frac{N_{\rm m}}{N_{\rm e}}b_{\rm e}$ and $B_{\rm m}=\frac{N_{\rm e}}{N_{\rm m}}b_{\rm m}$ respectively, for $b_{\rm e,m}$ integer classes. 

As an example, let us once again consider Maxwell theory on $\overline{\mathbb{CP}}^2$, where we gauge a $\mathbb{Z}_2\times\mathbb{Z}_3$ subgroup of the one-form symmetry. We pick background fields $B_{\rm e}=\frac{3}{2} b_{\rm e}$ and $B_{\rm m}=\frac{2}{3} b_{\rm m}$, and thus we get
\be
Z \genfrac[]{0 pt}{0}{\frac{3}{2}k}{\frac{2}{3}l} &=& ({\rm Im}\,\tau)^{-\frac{1}{2}} \sum_{ m\in \mathbb{Z}}e^{\pi {\rm i}\tau \left( m + \frac{3}{2} k \right)^2}e^{2 \pi {\rm i} \left( m + \frac{3}{2} k \right) \frac{2}{3}l} \, ,
\ee
which is invariant under $k \to k +2 \mathbb{Z}$ and $l \to l +3 \mathbb{Z}$.\footnote{These are the large gauge transformation of the background gauge fields. They are related to the arbitrariness in the choice of integral lifts of these fields.} Hence, the gauging can be performed by summing over $k=0,1$ and $l=0,1,2$, leading to
\be
\sum_{k,l} Z \genfrac[]{0 pt}{0}{\frac{3}{2}k}{\frac{2}{3}l}[\tau] &=& \frac{9}{2} Z \genfrac[]{0 pt}{0}{0}{0}[\tfrac{9}{4} \tau] \, ,
\ee
which is compatible with \eqref{gauged_maxwell}, using the normalization in \eqref{eq:discretebackground}. 

With this analysis, it is now easy to see how one can define a gauging on the partial theory. For example, one can consider $M_{E_8} \# \overline{M}_{E_8}$ and perform an electric gauging $G^{\rm sd}_{N_{\rm e}}$ on the self-dual sector only
\be
Z_{G^{\rm sd}_{N_{\rm e}}} [\tau]= N_{\rm e} \sum_{\bf n} Z_{M_{E_8}} \genfrac[]{0 pt}{0}{\frac{\bf n}{N}}{0}[\tau] \ \frac{Z_{\overline{M}_{E_8}} [\tau]}{ Z_{\mathcal{S}^4}[\tau]} =\frac{ Z_{M_{E_8}}[\tfrac{\tau}{N_{\rm e}^2}]Z_{\overline{M}_{E_8}} [\tau]}{Z_{\mathcal{S}^4}[\tau]} \, .
\ee
With this notion of ``partial'' gauging, we can repeat the analysis of the previous section and define a ``partial'' non-invertible symmetry by combining a partial gauging with a partial symmetry.
Consider, for example, a theory at coupling $\tau = N^2 {\rm i}$, whose partition function admits the splitting 
\be
Z_{\mathcal{M} \# \mathcal{N}} [N^2 {\rm i}]&=& \frac{Z_{\mathcal{M}} [N^2 {\rm i}] Z_{\mathcal{N}} [N^2 {\rm i}]}{Z_{\mathcal{S}^4}[N^2{\rm i}]} \, .
\ee
One can now perform a (electric) $G_N$ gauging only on the partial theory defined on $\mathcal{M}$, followed by a partial $S$-duality and finally by a (magnetic) $G_{\frac{1}{N}}$ gauging, such that the coupling goes back to the starting value $G_{\frac{1}{N}} S G_N (N^2 {\rm i}) = N^2 {\rm i}$. This duality-symmetry operation is now non-invertible and it is ``partial'', in the sense that it only acts on the self-dual fluxes of the theory.

\subsection{Zeros vs anomalies}\label{Sec:Leech}

There is a fundamental difference between the invertible $SL(2,\mathbb{Z})$ duality of Section \ref{Sec:Inv} and its non-invertible $SL(2,\mathbb{Q})$ (or $SL(2,\mathbb{R})$) generalization: The latter does not correspond to a ``gauge'' symmetry in the parameter space $\mathbb{H}$. In other words, configurations connected by $SL(2,\mathbb{Q(R)})$ transformations do not have to be regarded physically indistinguishable.\footnote{Note that, if this were the case, $Z$ would have been a section of a generalized equivariant bundle which, if non-trivial, would have implied that $Z$ is identically zero. This is due to the fact that almost any point in $\mathbb{H}$ can be reached from any other point by a $SL(2,\mathbb{Q})$ transformation.} This is because, as we have reviewed above, the non-invertible defect operators remove some lines from the spectrum.

However, every point in $\mathbb{H}$ now (and not just the orbifold ones as before) is fixed by a non-trivial subgroup of $PSL(2,\mathbb{Q(R)})$: The subgroup is isomorphic to $PSO(2,\mathbb{Q(R)})$ and depends on the point. Analogously to the $\mathbb{Z}_2$ and $\mathbb{Z}_3$ subgroups of $PSL(2,\mathbb{Z})$ of the invertible case, this stabilizer group would be associated to a now non-invertible duality-symmetry of Maxwell theory corresponding to almost any value of the coupling in $\mathbb{H}$ (or literally at any coupling if we consider the $\mathbb{R}$ version) \cite{Niro:2022ctq}. Such symmetries may have a mixed 't Hooft anomaly with gravity, and it is tempting to attribute a zero of the partition function at a smooth point $\tau_*$ to this anomaly. The anomaly is, of course, an intrinsic feature of the non-invertible symmetry defined at a given $\tau_*$.

One concrete consequence of this anomaly would be the following. Consider a matrix $M\in PSL(2,\mathbb{Q(R)})$ such that $M(\tau_*)={\rm i}$. Then the transformation $M^{-1}SM$ generates a $\mathbb{Z}_2$ subgroup of the stabilizer group of the point $\tau_*$. The anomaly thus implies that
\be\label{AnomPhasZ2NonInv}
Z[\tau_*] = - Z[M^{-1}SM(\tau_*)]\,,
\ee
Provided that $Z$ does not vanish at the triality point, a similar discussion can be carried over for a $\mathbb{Z}_3$ subgroup of the stabilizer group of $\tau_*$, using a matrix $N\in PSL(2,\mathbb{Q(R)})$ such that $N(\tau_*)=\rho$.

Let us make a remark here. Since $M$ or $N$ necessarily contain a (possibly infinite) number of gaugings of one-form symmetries, one would naively ascribe the above anomaly to a mixed one between the one-form symmetries and gravity. However, this cannot be the case: A mixed anomaly between, say, a $\mathbb{Z}_n\times\mathbb{Z}_m$ one-form symmetry group and gravity, if present, would be independent of the value of $\tau$ and, starting from the theory at any value of $\tau$ such that $Z[\tau]\neq0$, it would kill $Z[\tfrac{N_{\rm m}^2}{N_{\rm e}^2}\tau]$ by the gauging procedure for any $N_{\rm e}$ multiple of $n$ and $N_{\rm m}$ multiple of $m$. This would imply that the partition function will vanish identically. A similar argument can also be used to exclude the presence of mixed anomalies between the one-form symmetries and the invertible zero-form symmetries arising at the orbifold points: If present, they would create many more zeros for $Z$ than e.g.~those allowed in the split case by formula \eqref{ZeroHolMod}.

We do not have a direct way to assess whether anomalous phases such as the one in \eqref{AnomPhasZ2NonInv} are present or not. Hence, we rely on the vanishing of the partition function as a criterion to detect anomalies. Therefore, in the following, rather than aiming to be exhaustive, we limit ourselves to presenting one concrete example where the partition function of Maxwell theory has a single zero of order one at a smooth point of the conformal manifold.

Start with the connected sum of $24$ copies of the product manifold $S^2\times S^2$. As we have seen in Section \ref{new symmetry}, we can isomorphically map its intersection form to that of the manifold $M_{E_8}^{\#3}\#\overline{M}_{E_8}^{\#3}$. By choosing an extremal metric $g_{E_8}$, for which $E_8^{\oplus 3}$ is the lattice of self-dual fluxes, we have seen that $\tilde{Z}[g_{E_8};\tau,\bar{\tau}]=E_4[\tau]^3\bar{E}_4[\bar{\tau}]^3$, which vanishes at the triality point as the effect of anomalous partial triality-symmetries. Remarkably, however, there exists another isomorphism of intersection forms, which instead takes us to $L\oplus \bar{L}$, where $L$ is the so-called Leech lattice, an even unimodular $24$-dimensional lattice, thus establishing the following diffeomorphism
\be
(S^2\times S^2)^{\#24}&\simeq & M_L\# \overline{M}_L\,,
\ee
with $M_L\# \overline{M}_L$ the smooth, simply-connected manifold associated to $Q_{L\oplus \bar{L}}$. By choosing an extremal metric $g_L$, for which $L$ is the lattice of self-dual fluxes, the normalized partition function reads
\be\label{ZLeech}
\tilde{Z}_{ M_L\# \overline{M}_L}[g_{L};\tau,\bar{\tau}]&=& \vartheta_L(\tau)\, \bar{\vartheta}_L(\bar{\tau}) \,,
\ee
where $\vartheta_L(\tau)$ is the generalized theta function of the Leech lattice, a holomorphic modular form of weight $12$, which can be written as\footnote{The case of the Leech lattice can easily be generalized to a whole class of examples, given by the so-called Niemeier lattices, i.e.~the $24$ positive-definite even unimodular lattices of rank $24$, whose generalized theta functions are parametrized by the Coxeter number $h$ of the lattice, and read $\vartheta_{N}(h;\tau)=\tfrac1{72}((42+h) E_4[\tau]^3+(30-h) E_6[\tau]^2)$. For the Leech lattice, $h=0$.}
\be\label{ThetaLeech}
\vartheta_L(\tau)&=&\frac{1}{12}\left(7 E_4[\tau]^3+5 E_6[\tau]^2\right)\,.
\ee
As is clear from \eqref{ZeroHolMod}, the normalized partition function \eqref{ZLeech} will now have only one simple zero at a smooth point $\tau_*^{(L)}$ of the conformal manifold. We estimated its approximate location to be
\be
\tau_*^{(L)}\approx 0.23510834+0.97196917\,{\rm i}\,,
\ee
which we truncated at eight decimal digits. Notice that this zero is likely lying on the lower edge of the fundamental domain, i.e.~$|\tau_*^{(L)}|=1$. 

Given what we said above, it is natural to suspect that this zero is related to the mixed anomaly with gravity of a \emph{partial non-invertible} symmetry of Maxwell theory at $\tau=\tau_*^{(L)}$, transforming solely the self-dual (or equivalently the antiself-dual) fluxes. It would be very interesting to better clarify the nature of such a symmetry and to verify the presence of its anomaly by a direct computation of the anomalous phase.

\section{Discussion}\label{Sec:Conclu}

In this paper, we have analyzed some of the intriguing phenomena that originate from considering free quantum field theories supported on non-trivial geometries. The case of interest was Maxwell theory on compact four-manifolds, but our results could be easily extended to other free theories that contain Maxwell as a subsector, such as $\mathcal{N}=4$ Abelian gauge theory in four dimensions. The latter can be obtained from dimensional reduction on a two-torus of the $(2,0)$ theory living on a single M5 brane in flat space. Hence, the electromagnetic-duality behavior of the four-dimensional gauge theory is inherited from the way the six-dimensional theory, including its fermionic sector, reacts to large diffeomorphisms of the torus.

We showed that, under certain precise conditions on both the topology and the metric of the four-manifold, the (suitably normalized) partition function of Maxwell theory undergoes a holomorphic factorization. We associated the different factors of the factorization to ``would be'' Maxwell theories on generally non-smooth four-manifolds. This observation led to the discovery of a novel type of symmetry, with a mixed internal/space-time nature, arising at special values of the coupling and acting only on a subsector of flux configurations. We uncovered its peculiar features as well as its mixed 't Hooft anomaly with gravity. The factorization of the partition function allowed us to control its vanishing even in the absence of anomalous phases for the ordinary duality-symmetries, and we leveraged this knowledge to argue for the existence of mixed 't Hooft anomalies between gravity and the recently-constructed non-invertible version of duality-symmetries.

It would be interesting to see how much of this analysis survives in the case of interacting theories sharing analogous modular features, such as $\mathcal{N}=4$ Super-Yang-Mills theories in four dimensions, whose duality behavior and non-invertible symmetries have been discussed e.g.~in \cite{Aharony:2013hda} and \cite{Kaidi:2022uux} respectively. This analysis might lead to novel identities between topological invariant quantities, such as Donaldson-Witten invariants (see \cite{Manschot:2023rdh} for a recent review). 

It would also be important to analyze how small perturbations away from extremal metrics modify our results and possibly explore the interplay between the softly broken partial symmetries and the addition of charged matter.\footnote{We thank both the anonymous referees for suggesting this direction.}

Another very intriguing direction of further study concerns the non-Lagrangian theory of a self-dual two-form in six dimensions, which, upon dimensional reduction on a two-torus, yields the Maxwell theory in four dimensions, giving a target-space origin to the electromagnetic duality \cite{Verlinde:1995mz}. By dimensionally reducing this theory on the compact four-manifold $\mathcal{M}$, instead, one gets the theory of two-dimensional compact chiral bosons with target space the intermediate Jacobian $H^2(\mathcal{M},\mathbb{R})/H^2(\mathcal{M},\mathbb{Z})$, where the role of space-time isometries and dualities is exchanged with respect to Maxwell theory (see \cite{Gadde:2013sca,Bashmakov:2023kwo,Chen:2023qnv} for examples in the supersymmetric case). It would be interesting to see whether such a unifying six-dimensional perspective sheds more light on the partial symmetries we have introduced here, perhaps by clarifying their effect on the spectrum of lines, and provides some additional insights on the mixed anomalies between gravity and the non-invertible duality-symmetries. In particular, the partial symmetries might be related to discrete isometries of the metric of the six-dimensional theory, which would give them a geometric origin.

One might also consider compactifying the six-dimensional theory on genus $g>1$ Riemann surfaces and study duality anomalies and possible partial duality-symmetries of the ensuing $U(1)^g$ four-dimensional gauge theory. In addition, one could analyze the effect on the four-dimensional duality structure of compactifications on non-trivially fibered Riemann surfaces.

Finally, it is known that the electromagnetic duality of Maxwell theory is plagued by a \emph{pure} 't Hooft anomaly too, which can be detected by promoting the gauge coupling to a space-time dependent complex parameter \cite{Hsieh:2019iba,Hsieh:2020jpj}. It would be important to establish whether the non-invertible duality-symmetries are also affected by this pure anomaly, which would represent a gravity-background independent obstruction to gauging them. 

\subsection*{Acknowledgements}

We would like to thank Chiara Altavista, Francesco Benini, Andrea Cipriani, Michele Del Zotto, Matteo Dell'Acqua, Lorenzo Di Pietro, I\~naki Garc\'ia-Etxebarria, Azeem Hasan, Cristoforo Iossa, Thomas Kragh, Joe Minahan, Marco Nardecchia, Elias Riedel Gårding, Filip Strakos, Valdo Tatischeff, and Gianluca Zoccarato for enlightening discussions. Special thanks to Francesco Benini and I\~naki Garc\'ia-Etxebarria for useful comments on the manuscript.

\paragraph{Funding information}
MT is supported by the ERC-COG grant NP-QFT No.~864583
``Non-perturbative dynamics of quantum fields: from new deconfined
phases of matter to quantum black holes'' and by the MUR-FARE2020 grant
No.~R20E8NR3HX ``The Emergence of Quantum Gravity from Strong Coupling
Dynamics''. The work of SM and DM is supported by the Simons Foundation (grant \#888984, Simons Collaboration on Global Categorical Symmetries). DM is also supported by the VR project grant No.~2023-05590..

\begin{appendix}
\numberwithin{equation}{section}

\section{The space of harmonic two-forms on four-manifolds}\label{MathToolkit}

After reviewing some basic definitions concerning the geometry of four-manifolds, this appendix is devoted to an in-depth discussion about the existence of metrics saturating the bound \eqref{Bound}. First, in Section \ref{Sec:IntForm}, we define the intersection form of a four-manifold, which characterizes its topology. Then, in Section \ref{Sec:ConnSum}, we review the operation of connected sum, as it will be our main tool in constructing smooth four-manifolds starting from ``easy" building blocks. In Section \ref{Sec:ClassTh}, we present a powerful classification theorem of simply-connected four-manifolds, and, in Section \ref{Sec:MCG}, review the concept of mapping class group. Finally, in Section \ref{Sec:ExtMet}, we talk about the Hodge star and the existence of ``extremal" metrics. This material is taken mainly from \cite{scorpan2005wild} (Chapters 3 to 5) and \cite{donaldson1997geometry}.

\subsection{The intersection form}\label{Sec:IntForm}

Let us start by discussing the intersection form, which contains much information on the topology of our four-manifold $\mathcal{M}$. The space $H^2(\mathcal{M},\mathbb{Z})$ is endowed with the cup product, defining a symmetric, non-degenerate pairing
\begin{equation}
	\begin{split}
		Q:H^2(\mathcal{M},\mathbb{Z})\times H^2(\mathcal{M},\mathbb{Z}) \to \mathbb{Z} \\
		(a,b) \to \int_{\mathcal{M}} a\cup b \ .
	\end{split}
\end{equation}
Since we assume $H^2(\mathcal{M},\mathbb{Z})$ to be a free module, we can view an element $a\in H^2(\mathcal{M},\mathbb{Z})$ as a closed two-form with integer periods (up to shifts by exact forms). Then, the intersection form is simply given by 
\begin{equation}
	Q(a,b) = \int_{\mathcal{M}} a \wedge b \ .
\end{equation}

Here are some useful facts and definitions.
\begin{itemize}
	\item We say that $Q$ is positive (negative) definite if $Q(a,a) > 0$ ($<0$) for any non-zero $a\in H^2(\mathcal{M},\mathbb{Z})$. We also say that it is even if $Q(a,a)$ is even for any $a$, otherwise we say that it is odd.
	\item The signature of $Q$ is defined as the difference between the dimensions of the maximal positive and maximal negative-definite subspace of $Q$. Concretely, we can always diagonalize $Q$ over the reals and consider the spaces of positive and negative eigenvalues. This induces a decomposition $H^2(\mathcal{M},\mathbb{R}) = H^2_+(\mathcal{M},\mathbb{R})\oplus H^2_-(\mathcal{M},\mathbb{R})$. Let $b_2^\pm(\mathcal{M}) = {\rm dim} \, H^2_\pm(\mathcal{M},\mathbb{R})$. Then the signature is $\sigma = b_2^+ - b_2^-$.
	\item The intersection form is $\mathbb{Z}$-bilinear, symmetric, and, as a consequence of Poincar\'e duality, unimodular (i.e.~$\mathrm{det}\,Q = \pm 1$). Moreover, if the manifold is spin, the intersection form is even (the converse holds if $H_1(\mathcal{M},\mathbb{Z})$ has no two-torsion, which is assumed throughout).
	\item If $\mathcal{M}$ and $\mathcal{N}$ are simply-connected smooth four-manifolds with isomorphic intersection forms (i.e.~$Q_\mathcal{M}$ to $Q_\mathcal{N}$ are connected by a change of basis over the integers), then $\mathcal{M}$ and $\mathcal{N}$ are homeomorphic (as we will see, this is a consequence of Freedman's classification theorem).
\end{itemize}

Let us now give some examples.
\begin{itemize}
	\item The manifold $S^2 \times S^2$ has exactly two non-trivial two-cycles given by the two $S^2$ factors. These do not self-intersect, but they intersect each other at a point. Hence, the intersection form is given (in this basis) by $Q_{S^2\times S^2}=H=\left[\begin{smallmatrix}
		0 & 1 \\
		1 & 0
	\end{smallmatrix}\right]$, the ``hyperbolic plane''. It is easy to see that this form is even and, since the manifold is simply-connected, this implies that the manifold is spin.
	\item The manifold $S^2 \times S^2$ can be thought of as the trivial bundle $S^2\times \mathbb{C}$, where we compactify each fiber with the point at infinity. We can play a similar trick with any line bundle over $S^2$. These are classified by the first Chern number $n\in\mathbb{Z}$, and for any integer, we can construct a compact manifold by compactifying the fibers. This way we obtain the so-called Hirzebruch surfaces, $\mathbb{F}_n$, which are nothing but $S^2$ bundles over $S^2$. It is easy to compute, e.g.~by toric methods, the intersection form of $\mathbb{F}_n$: $Q_{\mathbb{F}_n}=\left[\begin{smallmatrix}
		0 & 1 \\
		1 & n
	\end{smallmatrix}\right]$. Remarkably, it turns out that, for $n$ even, $Q_{\mathcal{M}_c} \cong H$, whereas, for $n$ odd, $Q_{\mathcal{M}_c}\cong \left[\begin{smallmatrix}
		1 & 0 \\
		0 & -1
	\end{smallmatrix}\right]$. Hence, the surfaces $\mathbb{F}_{\rm even}$ are all homeomorphic (in fact diffeomorphic) to the direct product $S^2\times S^2$, whereas all the $\mathbb{F}_{\rm odd}$ to the twisted bundle $S^2\tilde{\times} S^2$, which is non-spin.
\end{itemize}

\subsection{Connected sums}\label{Sec:ConnSum}

One very important operation on four-manifolds that we are going to use extensively is the connected sum. If we have two four-manifolds $\mathcal{M}$ and $\mathcal{N}$, their connected sum $\mathcal{M}\# \mathcal{N}$ is obtained as follows. First, we remove a small four-ball from both $\mathcal{M}$ and $\mathcal{N}$. In this way, they become manifolds with boundary given by $S^3$. Then, we embed a pair of cylinders $S^3\times [0,1]$ to glue together the two $S^3$ boundaries in the following manner: one of the cylinders is such that $S^3\times\{1\}$ coincides with the boundary of $\mathcal{M}$ and $S^3\times[0,1)$ is in the interior of $\mathcal{M}$, while the other cylinder is such that $S^3\times\{0\}$ coincides with the boundary of $\mathcal{N}$ and $S^3\times(0,1]$ is in the interior of $\mathcal{N}$. The connected sum is then obtained by identifying these two cylinders. The following are three important properties of connected sums.
\begin{itemize}
	\item Summing an $\mathcal{S}^4$ will not change the topology of the connected sum: $\mathcal{M}\# \mathcal{S}^4 \cong \mathcal{M}$.
	\item If we use the Mayer-Vietoris sequence on the open manifolds engineered to build the connected sum, we have $H^i(\mathcal{M}\# \mathcal{N} ,\mathbb{Z})=H^i(\mathcal{M} ,\mathbb{Z})\oplus H^i( \mathcal{N} ,\mathbb{Z})$ for $0<i<4$. As a consequence, we have that $\chi(\mathcal{M}\# \mathcal{N}) = \chi(\mathcal{M}) + \chi(\mathcal{N}) - 2$.
	\item If $\mathcal{M}$ and $\mathcal{N}$ have intersection form $Q_\mathcal{M}$ and $Q_\mathcal{N}$, their connected sum $\mathcal{M}\# \mathcal{N}$ has intersection form $Q_\mathcal{M}\oplus Q_\mathcal{N}$. Conversely, if $\mathcal{M}$ is simply-connected and $Q_\mathcal{M}$ splits as a direct sum $Q_1\oplus Q_2$, then there exist topological four-manifolds $\mathcal{N}_1$ and $\mathcal{N}_2$ such that $\mathcal{M}\cong \mathcal{N}_1\# \mathcal{N}_2$ (as we will see, this is a corollary of Freedman's classification theorem).
\end{itemize}

As a first example, consider $\mathbb{CP}^2 \# \overline{\mathbb{CP}}^2$, with the bar denoting reversed orientation. Since $H^2(\mathbb{CP}^2, \mathbb{Z})=\mathbb{Z}$, we have that the intersection form is given by a single integer. The generator of the second cohomology group of $\mathbb{CP}^2$ is given by the Poincaré dual to $\mathbb{CP}^1 \subset \mathbb{CP}^2$. Since two $\mathbb{CP}^1$s intersect generically in a point, we conclude that $Q_{\mathbb{CP}^2} = 1$. Reversing the orientation flips all the signs in the intersection form, leading to $Q_{\overline{\mathbb{CP}}^2} = -1$. Hence, $Q_{\mathbb{CP}^2 \# \overline{\mathbb{CP}}^2}=\left[\begin{smallmatrix}
	1 & 0 \\
	0 & -1
\end{smallmatrix}\right]$, which implies that $\mathbb{CP}^2 \# \overline{\mathbb{CP}}^2\cong S^2\tilde{\times} S^2$. Another example, particularly important for the purposes of this paper, is the connected sum $(S^2\times S^2)^{\#8}$. Remarkably, its intersection form can be diagonalized over the integers into the form $[+E_8] \oplus [-E_8]$, where 
\begin{equation}\label{E8matrix}
	[+E_8] = \begin{bmatrix}
		2 & 1 &  &  &  &  &  &  \\
		1 & 2 & 1 &  &  &  &  &  \\
		& 1 & 2 & 1 &  &  &  &  \\
		&  & 1 & 2 & 1 &  &  &  \\
		&  &  & 1 & 2 & 1 &  & 1 \\
		&  &  &  & 1 & 2 & 1 &  \\
		&  &  &  &  & 1 & 2 &  \\
		&  &  &  & 1 &  &  & 2 
	\end{bmatrix} \ .
\end{equation}
Therefore we conclude that $(S^2\times S^2)^{\#8}\cong M_{E_8}\#\overline{M}_{E_8}$, where $M_{E_8}$ is a topological space with intersection form \eqref{E8matrix}. As we will see, by Donaldson's theorem, $M_{E_8}$ does not admit any smooth structure. However, the connected sum $M_{E_8}\#\overline{M}_{E_8}$, being homeomorphic to a connected sum of spheres, is smooth. Notice also that the $E_8$-form is positive-definite and even. This implies that $ (S^2\times S^2)^{\# 8} $ is a spin manifold.

\subsection{Classification theorems}\label{Sec:ClassTh}

Let us now present some very powerful classification theorems.
\begin{Theorem}[Serre's Classification]
	Let $Q:\mathbb{Z}\times\mathbb{Z}\to\mathbb{Z}$ be a symmetric, bilinear, and unimodular form. 
	\begin{itemize}
		\item If Q is indefinite and odd, then in a suitable basis it can be written as $$Q= [+1]^{\oplus m} \oplus [-1]^{\oplus n} \ .$$
		\item  If Q is indefinite and even, then in a suitable basis it can be written as $$ Q=  [\pm E_8]^{\oplus m}  \oplus H^{\oplus n} \ .$$
	\end{itemize}
\end{Theorem}
\begin{Theorem}[Freedman's Classification]
	For any integral, symmetric, unimodular form $Q$, there is a closed, simply-connected topological four-manifold that has $Q$ as its intersection form.
	\begin{itemize}
		\item If $Q$ is even, there is exactly one such manifold.
		\item If $Q$ is odd, there are exactly two such manifolds, and at least one of them does not admit any smooth structure.
	\end{itemize}
	In particular, if $\mathcal{M}$ and $\mathcal{N}$ are closed, simply-connected, smooth four-manifolds with isomorphic intersection forms, then they are homeomorphic.
\end{Theorem}
\begin{Theorem}[Donaldson]
	The only positive (negative) definite intersection forms of smooth manifolds are given by $[+1]^{\oplus m}$ ($[-1]^{\oplus n}$).
\end{Theorem}

From all these theorems and the fact that $(S^2\times S^2)^{\#8}\cong M_{E_8}\#\overline{M}_{E_8}$, we can extract the following important statement.

\begin{Corollary}
	Every smooth, simply-connected four-manifold not homeomorphic to $\mathcal{S}^4$ is homeomorphic to: 
	\begin{itemize}
		\item $ (\mathbb{CP}^2)^{\# m} \# (\overline{\mathbb{CP}}^2)^{\# n}$ if $Q$ is odd,
		\item $M_{E_8}^{\# m}\# (S^2\times S^2)^{\# n}$ or $\overline{M}_{E_8}^{\# m}\# (S^2\times S^2)^{\# n}$ if $Q$ is even.
	\end{itemize}
\end{Corollary}
Notice that, in general, there are constraints on how many $S^2\times S^2$ summands are allowed for every $M_{E_8}$ (or $\overline{M}_{E_8}$) summand such that the whole connected sum is smooth (for instance, $M_{E_8}^{\# 4}\# (S^2\times S^2)^{\# 5}$ does not admit a smooth structure \cite{furuta}, but $\overline{M}_{E_8}^{\# 2}\# (S^2\times S^2)^{\# 3}$ is smooth, being it homeomorphic to the K3 manifold). The precise number of how many $S^2\times S^2$ you can have for each $M_{E_8}$ is still an open problem in mathematics. Indeed, we have the following
\begin{conj}[11/8]
	Every smooth four-manifold $\mathcal{M}$ with even intersection form must have $b_2\geq \frac{11}{8} |\sigma|$.
\end{conj}
This means that we should have at least 3 $H$'s for each pair of $E_8$'s in $Q_\mathcal{M}$, namely that the most general simply-connected, smooth, spin four-manifold (not homeomorphic to $\mathcal{S}^4$) is homeomorphic to connected sums of K3 manifolds and $S^2\times S^2$'s. An important result in this direction, proven by Furuta in \cite{Furuta2004MONOPOLEEA}, is given by
\begin{Theorem}
	[10/8]
	Every smooth four-manifold with even intersection form must have $b_2\geq \frac{10}{8}|\sigma| + 2$  (i.e.~at least 2 $H$'s for every pair of $E_8$'s).
\end{Theorem}

\subsection{The mapping class group}\label{Sec:MCG}

Let us spend a few more words on Freedman's classification theorem. 

Consider two topological, simply-connected, four-manifolds $\mathcal{M}$ and $\mathcal{N}$. Suppose that there is a homeomorphism $f: M \to N$. Then, we have an isomorphism of cohomology groups $f^*: H^2(\mathcal{N},\mathbb{Z})\to H^2(\mathcal{M},\mathbb{Z})$. Hence, by identifying $H^2(\mathcal{M},\mathbb{Z})$ and $H^2(\mathcal{N},\mathbb{Z})$ with $\mathbb{Z}^{b_2}$, we see that performing a homeomorphism between two homeomorphic manifolds is equivalent to a change of basis of the second cohomology group, which depends only on the homotopy class of $f$. At the level of the intersection form, we have $Q_\mathcal{N} = L_f^T Q_\mathcal{M} L_f$, with $L_f\in GL(b_2,\mathbb{Z})$ being the matrix of change of basis induced by $f$ (i.e.~the two intersection forms are isomorphic).

Now, given two topological, simply-connected, four-manifolds with isomorphic intersection forms, can one always find a homeomorphism between them? The answer is in Freedman's classification theorem and it is ``almost". If $Q$ is even, then the answer is yes: Up to homeomorphisms there is a unique topological, simply-connected four-manifold with intersection form $Q$. However, when $Q$ is odd, we have exactly two. These are distinguished by the so-called Kirby–Siebenmann class, which is a class in $H^4(\mathcal{M},\mathbb{Z}_2)$, giving an obstruction for $\mathcal{M}$ to be smooth. In particular, the one with a non-trivial Kirby–Siebenmann class admits no smooth structure, while the one with trivial Kirby–Siebenmann class is smooth (being it homeomorphic to a connected sum of $\mathbb{CP}^2$'s and $\overline{\mathbb{CP}}^2$'s).

Therefore, given $Q$, provided that it is isomorphic to the intersection form of a smooth four-manifold, it is always realized as the intersection form of a smooth four-manifold.

So far we have focused on homeomorphisms between different topological four-manifolds (i.e. $\mathrm{Homeo}(\mathcal{M},\mathcal{N})$). Let us now focus on $\mathrm{Homeo}(\mathcal{M}, \mathcal{M})$. As before, we have that a homeomorphism $f:\mathcal{M}\to \mathcal{M}$ can be interpreted as a change of basis in $H^2(\mathcal{M},\mathbb{Z})$. However, this time, the intersection form is preserved. Hence, $L_f \in O(Q, \mathbb{Z})\equiv\{ L\in GL(b_2,\mathbb{Z}) \ | \ L^T Q L = Q\}$. In particular, a result by Quinn \cite{Quinn} states that we have an isomorphism
\begin{equation}
	\pi_0 \mathrm{Homeo}(\mathcal{M}) \cong O(Q,\mathbb{Z}) \ ,
\end{equation}
where $\pi_0 \mathrm{Homeo}(\mathcal{M})$ is by definition the topological mapping class group (the $\pi_0$ is due to the fact that we are only interested in the homotopy class of $f$).

So far, we have been concerned with topological manifolds and homeomorphisms. But, in applications, we are interested in smooth four-manifolds. Hence, we now look at how the above picture changes if we consider $\pi_0 \mathrm{Diff}(\mathcal{M})$ (i.e.~diffeomorphisms). 
Since every diffeomorphism is a particular case of homeomorphism, clearly we have a map
\begin{equation}
	\pi_0 \mathrm{Diff}(\mathcal{M}) \to O(Q,\mathbb{Z}) \ .
\end{equation}
However, contrary to the previous case, this map is in general guaranteed to be neither injective nor surjective. Nevertheless, there is an important result by Wall \cite{Wall}, which states
\begin{Theorem}
	Let $\mathcal{N}$ be a closed, oriented, simply-connected four-manifold, and suppose either that $Q_\mathcal{N}$ is indefinite or the rank of $H^2(\mathcal{N},\mathbb{Z})$ is at most 8. Then, if $\mathcal{M}$ is diffeomorphic to the connected sum $ \mathcal{N} \# (S^2\times S^2)$, the map $\pi_0 \mathrm{Diff}(\mathcal{M}) \to O(Q,\mathbb{Z})$ is surjective.
\end{Theorem}
Since most of the manifolds we consider are of this form, we will often assume surjectivity. With a slight abuse of terminology, we will refer to $O(Q_\mathcal{M},\mathbb{Z})$ as the mapping class group of $\mathcal{M}$.

\subsection{Metrics on harmonic forms}\label{Sec:ExtMet}

So far we have talked extensively about the topology of smooth four-manifolds, and we have been able to classify the simply-connected ones in terms of the intersection form. Now we shift our focus onto the second main character of our discussion, the Hodge-star operator.

If we endow our smooth manifold $\mathcal{M}$ with a Riemannian metric $g$, we can define the Hodge-star operator between two differential forms of the same degree as 
\begin{equation}
	\alpha \wedge\star \beta = \langle \alpha, \beta \rangle_g \,d\mathrm{vol} \ ,
\end{equation}
where $\langle\cdot,\cdot\rangle_g$ is the scalar product induced by the metric on differential forms, and $d\mathrm{vol}$ is the volume form of the metric.

On a four-manifold, the Hodge-star operator takes two-forms into two-forms. Moreover $\star\star = 1$ on two-forms. This means that we have eigenspaces with eigenvalues $\pm 1$. The elements of these eigenspaces are called self-dual and antiself-dual forms.
Given the rank-6 vector bundle of two-forms $\Lambda^{(2)}$, we can decompose it into the direct sum of two rank-3 vector bundles of self-dual and antiself-dual forms
\begin{equation}\label{sub-bundles}
	\Lambda^{(2)} = \Lambda^{(2)}_+\oplus\Lambda^{(2)}_- \ ,
\end{equation}
where $\alpha \in \Lambda^{(2)}_\pm$ is such that $\star \alpha = \pm \alpha$. This is obvious for any two-form $\alpha$ can be decomposed as $\alpha = \alpha^+ + \alpha^-$, where $\alpha^\pm = \tfrac{1}{2}(1\pm\star)\alpha$.

Since the Hodge-star operator on two-forms of four-manifolds depends only on the conformal class of the metric, we also have that the decomposition in self-dual and antiself-dual forms depends only on the conformal class of the metric. This point of view can be reversed. Indeed, we can regard any conformal structure for our manifold as being defined by the sub-bundle decomposition \eqref{sub-bundles}.

Let us now discuss the action of the Hodge star on $H^2(\mathcal{M},\mathbb{R})$. By the Hodge decomposition theorem, any element $\omega\in H^2(\mathcal{M},\mathbb{R})$ admits a unique representative which is harmonic (meaning that $d\omega = d\star\omega = 0$). The action of the Hodge-star operator sends harmonic forms into harmonic forms. Hence, we have a decomposition 
\begin{equation}
	H^2(\mathcal{M},\mathbb{R}) = H^2_+(\mathcal{M},\mathbb{R})\oplus H^2_- (\mathcal{M},\mathbb{R})\ .
\end{equation}
Clearly we have that an element $\alpha \in H^2_+(\mathcal{M},\mathbb{R})$ is such that
\begin{equation}
	\int_\mathcal{M} \alpha \wedge \alpha = \int_\mathcal{M} \alpha \wedge \star \alpha = \int_\mathcal{M} \langle \alpha,\alpha\rangle d\mathrm{vol} \geq 0 \ .
\end{equation}
Hence, the space $H^2_+(\mathcal{M},\mathbb{R})$ is a maximal positive-definite subspace of the intersection form (the same discussion holds for $H^2_-(\mathcal{M},\mathbb{R})$, which is a maximal negative-definite subspace).

\subsection{``Extremal'' metrics}

In this section we will discuss the existence of a class of metrics that saturate the bound \eqref{Bound}.\footnote{We thank Thomas Kragh for insightful discussions on this topic.}

Let us start by stating the problem that we want to solve. Suppose that we have chosen a topology for our four-manifold by fixing the intersection form $Q$. Over the reals we can always decompose the space of harmonic two-forms into a maximal positive and a maximal negative-definite subspace under $Q$, i.e.~$H^2(\mathcal{M},\mathbb{R})=H^2_+(\mathcal{M},\mathbb{R})\oplus H^2_-(\mathcal{M},\mathbb{R})$. Can we find a conformal class of metrics for which all the elements of $H^2_{+(-)}(\mathcal{M},\mathbb{R})$ are represented by (anti)self-dual harmonic two-forms?

As discussed in the previous section, a conformal class of metrics is equivalent to choosing a rank-3 vector bundle $\Lambda^{(2)}_+$ of self-dual differential two-forms. Hence, one is led to find a bundle $\Lambda^{(2)}_+$ in such a way that the space of its global sections contains one representative for each element in $H^2_+(\mathcal{M},\mathbb{R})$. These representatives are going to be the self-dual harmonic two-forms (which then determine the antiself-dual harmonic forms).

Here are some results on four-manifolds related to the existence of such metrics. We denote by $\mathrm{Gr}^+(\mathcal{M})$ the space of all possible maximal positive-definite subspaces of $H^2(\mathcal{M},\mathbb{R})$, and by $\mathrm{Met}(\mathcal{M})$ the space of all possible metrics on $\mathcal{M}$. Then, we can consider the so-called period map
\begin{equation}
	\Pi:    \mathrm{Met}(\mathcal{M}) \to \mathrm{Gr}^+(\mathcal{M}) \ .
\end{equation}
In \cite{Katz2003FourmanifoldSA} it is conjectured that the period map is surjective. This would imply the existence of the metrics we are seeking. Although this is still an open conjecture, in recent times a useful result in this direction has been proven in \cite{scaduto2023metricstretchingperiodmap}:
\begin{Theorem}\label{ScadutoTh}
	For a smooth, closed, connected, oriented four-manifold, $\Pi$ has dense image. Moreover, if $b_2^+ = 1$ (or equivalently, $b_2^-=1$), then the period map is surjective.
\end{Theorem}

To summarize, we now have the following picture:
\begin{itemize}
	\item If $[+E_8]\subset Q$, and $Q$ is the intersection form of a smooth four-manifold, then it is conjectured that there exists a metric in which the $[E_8]$ factor is self-dual.
	\item If $[+E_8] \subset Q$, and $Q$ is the intersection form of a smooth four-manifold with $b_2^- = 1$, then there exists a metric in which the $[E_8]$ factor is self-dual.
\end{itemize}

Let us give a couple of examples: 
\begin{itemize}
	\item $Q = [+E_8]\oplus[-E_8]$ is the intersection form of a smooth four-manifold homeomorphic to $(S^2\times S^2)^{\#8}$. If the conjecture holds true, it admits a class of metrics where the $[E_8]$ subspace is self-dual.
	\item $Q = [+E_8] \oplus [-1]$ is the intersection form of a smooth four-manifold homeomorphic to $(\mathbb{CP}^2)^{\# 8}\#\overline{\mathbb{CP}}^2 $, which does admit a class of metrics where the $[E_8]$ subspace is self-dual.
\end{itemize}

\end{appendix}





\bibliography{biblio.bib}


\end{document}